# Understanding and Designing Phase Change Materials: Insights from Atom Probe Tomography

*Jan Köttgen[1], Nils von den Driesch[2], Alexander Pawlis[2], Matthias Wuttig[1, 2, *]*

**J. Köttgen, Prof. M. Wuttig**

[1] Institute of Physics IA, RWTH Aachen University, 52074 Aachen, Germany

* E-Mail: wuttig@physik.rwth-aachen.de

**Dr. N. von den Driesch, Dr. A. Pawlis, Prof. M. Wuttig**

[2] Peter-Grünberg-Institute – JARA-Institute Energy Efficient Information Technology (PGI-10) Wilhelm-Johnen-Straße, 52428 Jülich, Germany

## Abstract

Phase Change Materials (PCMs) can be rapidly and reversibly switched between their amorphous and crystalline state; a transition which is accompanied by a pronounced change of optoelectronic properties. Here progress is reviewed to explain these property changes, focusing on advances by atom probe tomography (APT). This technique classifies bonding by providing two crucial bonding descriptors. Most important is the Probability of Multiple Events (PME), which is related to the likelihood that more than one ion is dislodged per successful laser pulse in laser assisted field evaporation. Crystalline PCMs are characterized by a PME above 55%, not found for metals or iono-covalent solids. This confirms that crystalline PCMs employ a unique bonding mechanism coined metavalent bonding (MVB). While crystalline PCMs employ MVB, amorphous PCMs behave as covalent solids characterized by a much lower PME. PCMs thus change their bonding upon crystallization, consistent with quantum-chemical calculations of bonding. Crystalline solids with a high PME lie in a narrow conductivity range between metals and iono-covalent solids, indicative for a competition between electron localization and delocalization. A map quantifying chemical bonding locates metavalent solids in a region where approximately one electron is shared between adjacent atoms and bonding is not too ionic. This quantum chemical bonding map is now used to find and explain property trends relevant for PCMs in various application domains.

## Introduction

This review summarizes the present understanding of phase change materials (PCMs). Unfortunately, the term phase change material is not well defined. PCMs are presently used in two quite different application domains, in information storage and in thermal energy storage, where different material properties are relevant. Hence, it should be no surprise that different material classes fulfill the requirements of information and thermal energy storage. The present review focusses on PCMs for information storage and computing. The corresponding phase change materials show a remarkable combination of properties [1] [2, 3]. Upon crystallization they change their optical and electrical properties considerably. The electrical resistivity, for example, often decreases upon crystallization by 4-6 orders of magnitude, while their optical properties also alter significantly [3] . These property changes can be achieved on short time scales, i.e. the time-limiting crystallization process can be realized on a sub-µs-time scale [4] or even sub-ns-time scale [5, 6]. This combination of properties makes PCMs attractive as a 'nanoswitch', i.e., a small (nm-size) and fast (ns-time) switch. Such a switch can, and has been, employed in rewritable optical data storage [2] and non-volatile electronic memories [7, 8], in photonic switches [9], neuromorphic computing [10, 11] as well as reconfigurable metasurfaces [12]. For these applications it is crucial that the metastable phase, which is usually amorphous and the more stable crystalline phase can coexist at the operation temperature of the device. This requires long time stability of the amorphous phase at this temperature. At the same time, it excludes using thermochromic materials like $VO_2$, which undergoes a phase transition with significant property changes, but only offers a volatile transition governed by sample temperature.

### A brief overview of the history of phase change materials

In the following, we will sketch the history of the discovery of PCMs and the subsequent development of concepts to understand their unconventional properties. Initially, these approaches focused on a determination of the atomic arrangement, assuming that this arrangement is closely related to the material properties. Recently, however, the focus has shifted to a classification of chemical bonding. After reviewing these developments, the resulting ideas to understand, explain and design phase change materials will be presented.

Research on chalcogenides and their potential for information storage gained considerable momentum already in 1968 when S.R. Ovshinsky [13] discovered reversible switching phenomena in amorphous films. Shortly afterwards, he and his coworkers [14] also identified light-induced switching processes. The application potential of PCMs for optical data storage was realized as early as 1986, when Chen and Rubin, working at the IBM Almaden Research Lab, showed that GeTe could crystallize in less than 2 µs and revealed significant changes of optical properties [15]. In 1988, the Japanese company Matsushita demonstrated superior performance for alloys along the pseudo binary line between GeTe and $Sb_2Te_3$ and shortly afterwards presented the first recorder for rewritable optical data storage [16]. In the following two decades three generations of rewritable storage media were developed from compact discs (CD) to digital versatile discs (DVD) and finally Blu-ray discs (BD). These different formats increased the storage density and improved data transfer rates [2]. In these early years, phase change materials were largely developed and improved by industry, primarily by the need to tailor phase change materials to decreasing laser wave lengths enabling higher storage capacities of the optical discs.

At the beginning of this millennia, it became clear that it would be very difficult to launch a 4th generation of successful rewritable storage media. Such a storage format met considerable technical challenges starting with the lack of a compact, inexpensive laser diode operating in the UV to phase change materials that would offer sufficient property contrast in this wave length range. At the same time, different companies realized that phase change materials could also be utilized in non-volatile electronic memories [7, 17, 18]. Furthermore, phase change materials were considered in novel photonic applications [9, 12]. These research initiatives coincided with a growing interest from academia, after it became clear that the pronounced property change upon crystallization is a rare feature in solids. Hence, the search for materials with particularly fast crystallization and distinct property contrast started. These activities were accompanied by attempts to unravel the origin of the property change.

## Attempts to understand the properties of PCMs: focus on atomic arrangement

Early attempts centered on an understanding of the atomic arrangement of the crystalline phase. Yamada and coworkers were the first to stress the octahedral or octahedral-like atomic arrangement of compounds along the pseudo binary line between GeTe and $Sb_2Te_3$ [16]. They also emphasized that these compounds frequently host high vacancy concentrations [19, 20]. In the following years, a number of phase change materials were identified, which differed significantly in stoichiometry from compounds along the pseudobinary-line mentioned above [21, 22]. This helped to realize that a wider class of materials offered a distinct property change upon crystallization. Furthermore, it became clear that the atomic arrangement, i.e. the short-range order of amorphous and crystalline phase change materials differs significantly [23]. This finding caught considerable attention and initiated a surge of research interest [24]. Kolobov and coworkers showed convincingly that the atomic arrangement of the amorphous and crystalline phase differed strongly in $Ge_2Sb_2Te_5$. They could demonstrate in particular that the next nearest neighbor distance in this compound *increased* (!) upon crystallization, even though this phase has a higher density. The key conclusion from this study, however, has been the claim that Ge changes from a tetrahedral arrangement in the amorphous state to a distorted octahedral arrangement in the crystalline phase [23]. A competing model has been developed by Huang and Robertson who argue that the main difference between amorphous and crystalline phase change materials like GeTe is the size of the Peierls distortion, i.e. the magnitude of the atomic arrangement away from the perfect octahedral symmetry [25]. Apparently, many phase change materials employ a crystal structure which can be attributed to a Peierls distortion. This distortion could thus be a characteristic pattern of PCMs [26-28]. Such a distortion has even been identified in liquid GeTe [27]. However, there are a number of compounds such as GeSe that can also be described as a Peierls distorted solid [29], yet GeSe lacks a large property contrast upon crystallization. Hence, the existence of a Peierls distortion in the crystalline phase is not sufficient to identify PCMs. On the other hand, the model based on tetrahedral Ge fails to explain how this mechanism can work in PCMs which do not contain Ge and where a tetrahedral arrangement of atoms is unlikely, such as elemental Sb [30]. Hence, it seems fair to say that past discussions of the atomic arrangement of crystalline PCMs have not yet identified a single characteristic pattern beyond doubt which prevails in all PCMs.

Later on, studies focused on the atomic arrangement in amorphous phase change materials [31-33]. While it is quite challenging to experimentally determine the atomic arrangement of amorphous materials with high precision, recent advances in the theoretical description have facilitated this task for amorphous materials. Molecular dynamics calculations enable the preparation of glassy solids/ undercooled liquids upon quenching liquid phase change materials. Subsequent studies have investigated structural relaxation in these glassy solids and their subsequent crystallization [34]. By now, these models have become so sophisticated and advanced that crucial insights on crystallization kinetics can be obtained on relevant length and time scales [31, 35-37]. Still, it seems daunting to predict distinct stoichiometry trends. This shortcoming impedes the rapid identification and design of new phase change materials. Hence, one can wonder if other concepts can be developed which help to understand, explain and tailor the unconventional property portfolio of PCMs.

## Chemical bonding as a concept to understand phase change materials

While most attempts to unravel the property portfolio of PCMs have focused on studies of the atomic arrangement, a few others have put chemical bonding into the spot light. Lucovsky and White already concluded in 1973 that crystalline GeTe employs resonant (or resonance) bonding [36]. Yet, they also argued that GeSe would utilize the same bonding mechanism, even though its crystal structure is considerably more distorted and GeSe does not provide the pronounced property contrast that characterizes PCMs. They attributed the underlying bonding pattern to resonant bonding [36] , a bonding mechanism suggested by Linus Pauling [38]. It has been employed e.g., to explain the atomic arrangement and properties of benzene. In 2010, the unusual properties of a number of crystalline

PCMs have also been attributed to an unconventional bonding mechanism [39]. In particular, a strong increase of the Born effective charge Z*, a measure of the chemical bond polarizability, was noted upon crystallization [40]. Furthermore, a pronounced increase of the optical dielectric constant $\varepsilon_\infty$ and a large decrease of the band gap $E_G$ upon crystallization was reported. Resonant / resonance bonding [41] was identified as the bonding mechanism responsible in the crystalline state [39]. On the contrary, the amorphous state of phase change materials has been attributed to covalent bonding. While this model, i.e., the change of bonding upon crystallization could convincingly explain the pronounced property change, invoking the concept of resonant bonding raised concerns.

If the unique properties of crystalline phase change materials are attributed to resonant bonding, one would expect that other solids which also utilize resonant bonding should show similar properties. However, this was not the case as shown by comparing the properties of benzene and graphene with properties characteristic for GeTe, SnTe and PbTe [42]. While the latter three chalcogenides reveal very high values of the optical dielectric constant $\varepsilon_\infty$ as well as the Grüneisen parameter of the transverse optical modes $\gamma_{TO}$, this is not the case for benzene or graphite. This pronounced difference casts serious doubt on the statement that crystalline phase change materials use the same bonding as graphite, or graphene. Subsequently, the question was answered as to whether crystalline PCMs have a unique portfolio of properties at all. This question was answered decisively in the affirmative [43].

Four different groups of solids could be classified based on their properties. These classes were closely related to four different bond types in solids. Besides covalent, ionic, and metallic bonds, a fourth strong bond was identified which apparently prevails in solids which are characterized by high values of the optical dielectric constant $\varepsilon_\infty$, a measure of the opto-electronic polarizability, large values of the Born effective charge Z*, a quantification of the chemical bond polarizability, as well as the Grüneisen parameter for the transverse optical mode $\gamma_{TO}$, a measure for the bond anharmonicity. This finding strongly suggests that a fourth strong bond type exists in solids. This unusual bond type has been coined metavalent bonding [42, 44]. To further substantiate this claim, we need quantitative descriptors to precisely characterize bonds in solids. Furthermore, we need to identify which material properties are closely related to chemical bonding. Both questions have lately been answered.

In recent years, several different program packages became available, which quantify chemical bonding in solids based on quantum chemical calculations [45-48]. These programs either use density- or orbital based approaches to derive these descriptors. While these programs provide a plethora of numbers, two quantities have proven particularly relevant to characterize chemical bonding; the number of electrons transferred (ET) to / from an atom and the number of electrons shared (ES) between adjacent atoms [49]. The number of electrons transferred is often still normalized by the oxidation state of the relevant atom [50]. Instead of the number of electrons shared (ES) between adjacent atoms, also the number of electron pairs formed, i.e., half of the number of electrons shared is specified. This latter number corresponds to the bond order. If these two quantities are used to span a map for simple solids, a map as shown in Figure 1 is obtained. Interestingly, in this map the different types of chemical bonds are well separated. While ionic bonds, as expected, are characterized by significant electron transfer, metallic bonds are characterized by the small number of electrons shared between adjacent atoms. This is a consequence of the large number of nearest neighbors which clearly exceeds the number of electrons these atoms have to form bonds. Most covalent elements like diamond or silicon share about two electrons (one electron pair), i.e., they form a single bond to a given neighbor. Interestingly, the map in Figure 1 shows a fourth region where solids are located with properties that are incompatible with ionic, metallic or covalent bonding. In this region crystalline solids like GeTe, $Sb_2Te_3$ or elemental Sb are located. This implies that crystalline phase change materials employ a peculiar bond type which differs from covalent, metallic or ionic bonding. This statement has been confirmed by the property classification mentioned above [51].

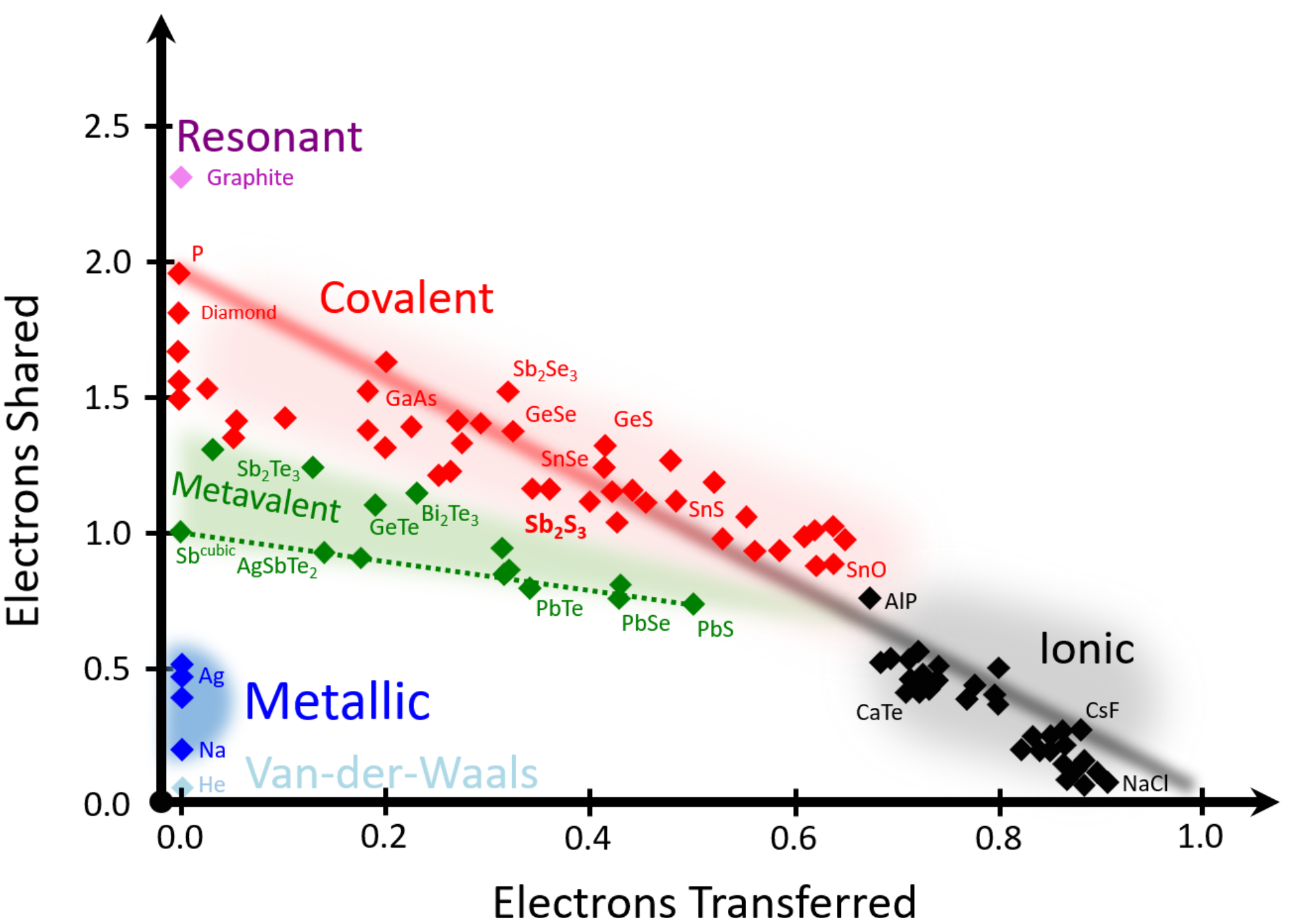

**Figure 1: 2D map classifying chemical bonding in crystalline solids.** The map is spanned by the number of electrons shared between adjacent atoms and the electron transfer normalized by the formal oxidation state. Different colors characterize different material properties and have been related to different types of bonds. The red–black line describes the transition from ideal covalent bonds to perfect ionic bonds. The dashed green line indicates metavalently bonded solids with perfect octahedral arrangement like cubic Sb, $AgSbTe_2$, and PbS, while distorted octahedrally coordinated structures are situated above it, characterized by a larger number of electrons shared. Redrawn under terms of the CC BY-NC 4.0 license [52], Copyright 2025.

Since metavalent solids possess a distinct property portfolio and occupy a well-defined region in the quantum chemical bonding map different from traditional bonding mechanisms (metallic, ionic and covalent) they apparently utilize a distinctively different bonding mechanism. Yet, it would be highly desirable to obtain further experimental evidence to distinguish bonding in solids. Ideally, such experiments should even help to understand and explain the different bonding mechanisms. Surprisingly, atom probe tomography turned out to be the technique that provides precisely this information.

## Atom Probe Tomography: a tool to distinguish bonding mechanisms in solids

Atom Probe Tomography (APT) enables the determination of the three-dimensional distribution of elements in a solid with near atomic resolution. In laser assisted field evaporation a high electrical field is applied to a tip shaped specimen. A short laser pulse of approximately 10 ps and sufficient energy is then adequate to overcome the activation barrier to dislodge ions. Two quantities help to classify the different bond types, the probability to form molecular ions (PMI) and the probability to observe multiple events (PME) upon a successful laser pulse. Both quantities are defined and depicted in the supplement (Figure S1). The PMI distinguishes the size of the ion, i.e., if a single atom or a molecular cluster is dislodged. If all ions are molecules, then the PMI corresponds to 100%. A related quantity is the mean fragment size, i.e., the number of atoms that form the ion (MFS). The second quantity, the PME, characterizes another propensity, i.e., the probability that more than a single ion is detected upon one laser pulse. High PMEs can also be a consequence of molecular dissociation or artefacts. In previous literature it has been discussed in detail how to distinguish the different origins of high PMEs [53]. For the solids marked in green in Figure 2, the high PME is neither caused by molecular dissociation nor an artefact. Instead, the high PME of these solids is an inherent material property. Figure 2 visualizes data for the PMI and the PME for a larger number of solids.

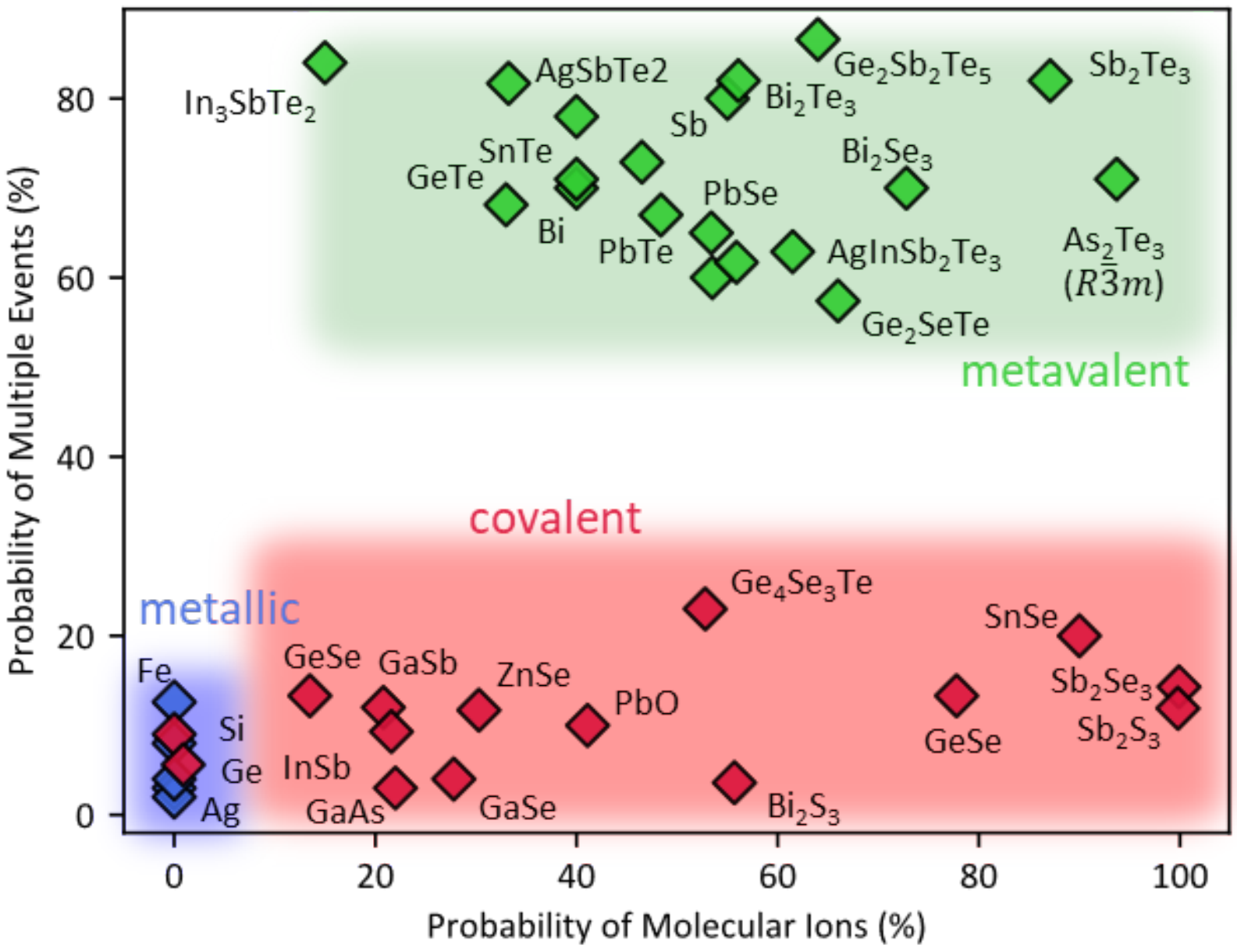

**Figure 2: Characteristic pattern of bond rupture for different crystalline solids.** Two quantities are depicted which enable a classification of bonding in various solids based on their bond rupture in atom probe tomography, the PME and the PMI. The latter quantity describes the probability that the ion formed is a molecule instead of an atom. Only metals show a vanishing PMI, i.e., for metals only atoms are dislodged. The PME characterizes the probability to form more than a single ion upon laser-assisted field evaporation. Usually, i.e., for metallic and covalent solids this PME is well below 25%. Metavalent solids, on the contrary, differ from all other solids depicted in Figure 1 by PME values above 55%. They hence possess a distinct bond rupture pattern which enables the fast identification of metavalent solids. A fully labeled version of the figure is provided in the supplement. Adapted under terms of the CC BY-NC-ND 4.0 license [53], Copyright 2025.

In contrast to metallic samples which evaporate one atom at a time, non-metallic materials can also evaporate as clusters of ions. The evaporation of molecular ions is common for iono-covalent solids. A second parameter is used to characterize the bond rupture, called the Probability of Multiple Events (PME). This number is determined as the ratio of the number of multiple events and the total number of events. The PME thus describes the probability to detect multiple ions evaporating upon the exposure of the tip to a single (successful) laser pulse. Interestingly, crystalline PCMs show a bond rupture which differs substantially from both metallic and covalent bonding. Unlike in metallic bonding, crystalline PCMs show a non-vanishing PMI; a feature which they have in common with covalent solids. Yet, both can be differentiated by their PME. Only crystalline PCMs show a high PME, neither observed in covalent nor metallic solids [54]. This is indicative for an unconventional bonding mechanism in crystalline PCMs, a finding which is sketched in Figure 2. This figure clearly reveals that APT can separate metallic, covalent and metavalent bonding. Particularly interesting is the clear difference in bond rupture between covalent and metavalent bonding. This is surprising since in the past, crystalline phase change materials have been discussed as narrow-gap semiconductors and covalent bonding has been implicitly assumed. Figure 2 shows unequivocally that crystalline PCMs employ a bond rupture which is incompatible with covalent bonding. At the same time, these crystalline PCMs also show a property portfolio which clearly differs from covalent solids [51]. This shows that both the unusual properties and the unconventional bond rupture indicators are apparently closely interwoven.

We can now return to the question if atom probe tomography can help to unravel how PCMs can be identified and understood. To this end, Figure 3 shows the bond rupture for a number of crystalline and amorphous / glassy solids [55]. While crystalline solids are depicted as diamonds, glasses are depicted as circles. Interestingly, the bond rupture hardly changes upon crystallization for those solids which employ covalent bonding in their crystalline state. This is shown for Ge, GeSe and $Sb_2S_3$ where the bond rupture hardly changes upon vitrification, i.e., the glassy and the crystalline solid both employ covalent bonding and reveal very similar bond rupture. This is very different for crystalline PCMs like $Ge_2Sb_2Te_5$, $In_3SbTe_2$, $GeSe_{0.5}Te_{0.5}$ and $GeSe_{0.25}Te_{0.75}$. For these crystalline solids, metavalent bonding is employed (as clearly shown by the unusual properties and the unconventional bond rupture). The corresponding amorphous phase (linked by an arrow) shows a very different bond rupture. These amorphous solids show a bond rupture which is characteristic for covalent bonding. Switching these solids between the amorphous and the crystalline state thus leads to a pronounced change in bond rupture and hence bonding. We can use this finding to explain how PCMs work and how to identify solids which possess this characteristic property portfolio.

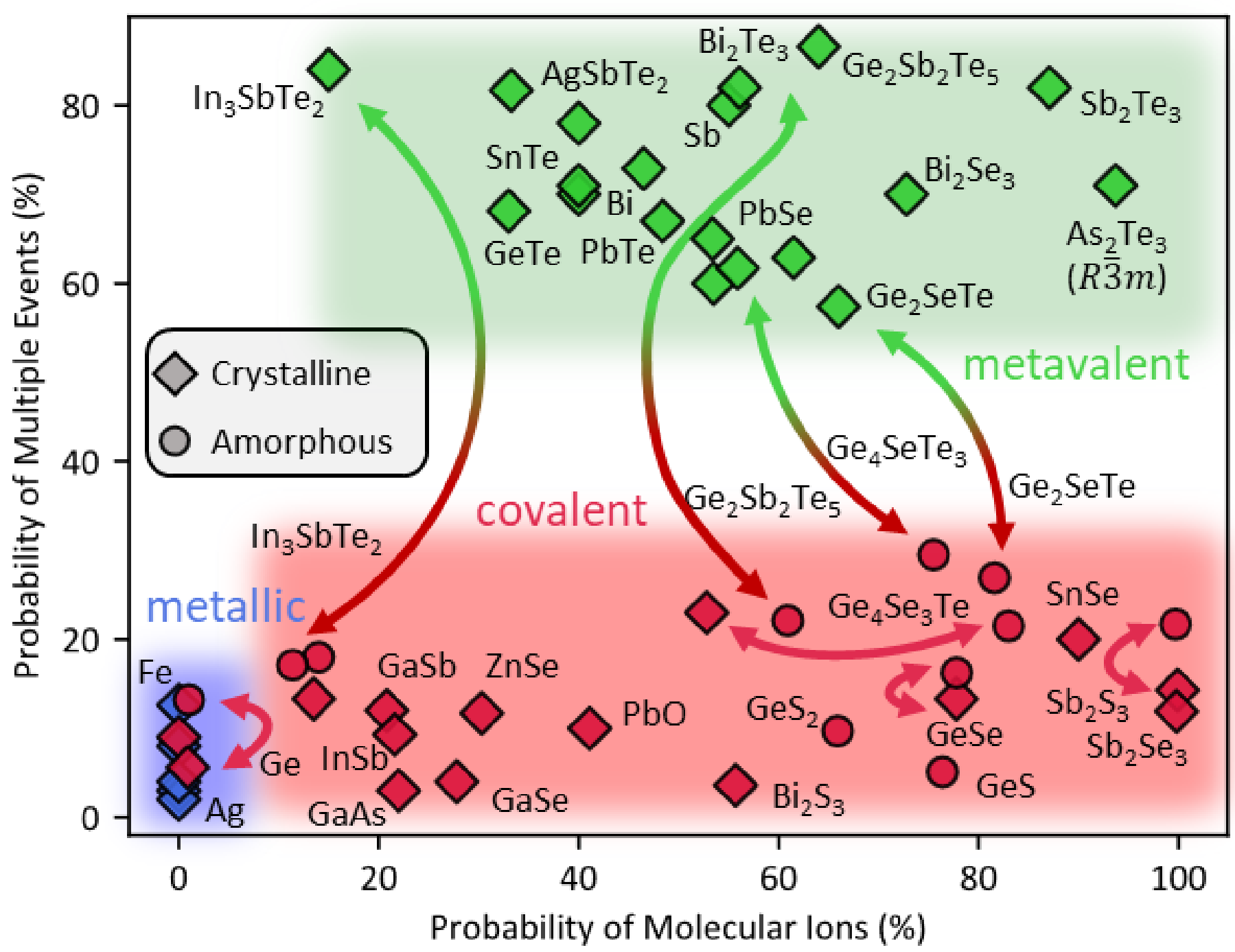

**Figure 3: Characteristic pattern of bond rupture for crystalline and amorphous/glassy solids.** The figure compares the bond rupture of a glassy solid (circle) with its corresponding crystalline phase (diamond). Covalent crystals like $Sb_2S_3$, GeSe or Ge hardly change their bond rupture upon vitrification [56]. Both their amorphous and crystalline phase show the characteristic bond rupture pattern of covalent solids. The situation is very different for metavalent crystals like $Ge_2Sb_2Te_5$, $In_3SbTe_2$, $GeSe_{0.5}Te_{0.5}$ and $GeSe_{0.25}Te_{0.75}$. For these solids a pronounced change of bond rupture is observed upon vitrification. While the amorphous phases of $Ge_2Sb_2Te_5$, $In_3SbTe_2$, $GeSe_{0.5}Te_{0.5}$ and $GeSe_{0.25}Te_{0.75}$ show the characteristic bond rupture of covalent solids, the corresponding crystals possess a bond rupture characteristic for metavalent solids. Hence, these solids change their bonding upon crystallization. Adapted under terms of the CC BY 4.0 license [55], Copyright 2025.

## Phase Change Materials as non-Zachariasen Glasses

Now, we can consider which solids exhibit a significant difference in properties between their crystalline and amorphous phases, particularly with respect to their opto-electronic characteristics. This is shown exemplarily in **Figure 4** for six different chalcogenides. Three chalcogenides—$Sb_2Te_3$, GeTe, and $GeSb_2Te_4$—show significant property changes upon crystallization, such as changes in the effective coordination number (ECON), the maximum of the imaginary part of the dielectric function $\varepsilon_{2,max}$, the optical dielectric constant $\varepsilon_\infty$ and the Born effective charge of the cations Z*, as displayed in Figure 4. Hence, these solids reveal the characteristic property portfolio of PCMs. These three crystalline chalcogenides show a bond rupture which is characteristic for metavalent solids, while their amorphous phases show the characteristic properties of covalent solids. In contrast, crystallization of covalent materials like $SiO_2$, $GeSe_2$ and GeSe results only in a subtle change of their properties. Compared with the changes of properties of the three PCMs, these differences are negligible. At the same time, these crystalline materials also show the characteristic properties of covalent solids, including the characteristic bond rupture depicted in Figure 2 (explicitly confirmed in Figure 2 for GeSe). Hence, there is compelling evidence that only metavalent solids show a pronounced change of properties upon crystallization, which is due to a pronounced change of bonding from covalent to metavalent upon crystallization.

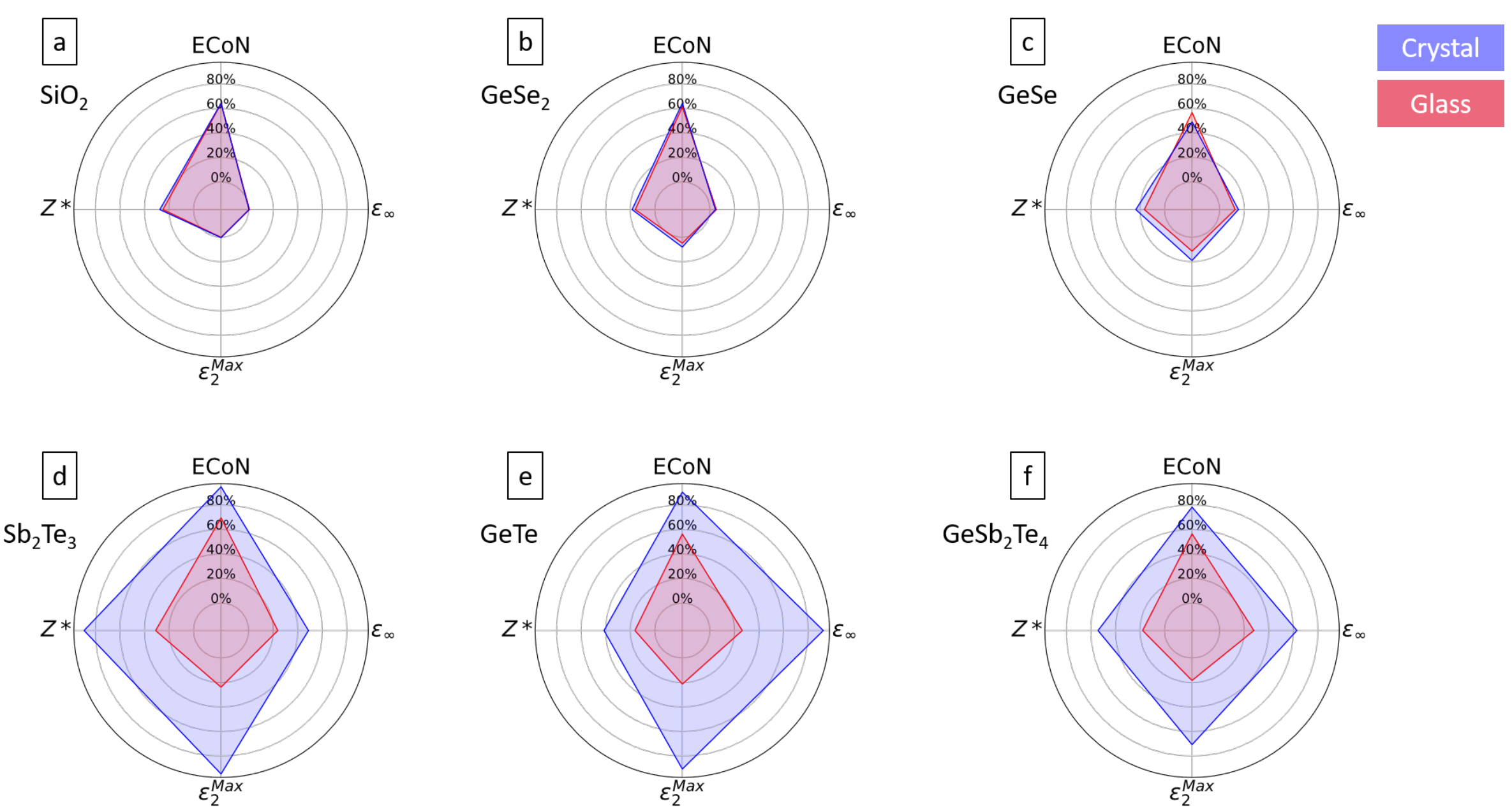

**Figure 4: Comparison of the properties of six different glasses (red) and their crystalline counterparts (blue).** Four different quantities are depicted to show the differences between glass and crystal; the effective coordination number (ECoN), the optical dielectric constant $\varepsilon_\infty$, the maximum height of the absorption peak $\varepsilon_{2max}$, as well as the (cation) Born effective charge $Z^*$. Pronounced changes of properties upon crystallization are only observed for the three non-Zachariasen glasses $Sb_2Te_3$, GeTe, and $GeSb_2Te_4$ (100% scale is ECoN = 6, $\varepsilon_\infty$ = 80, $\varepsilon_{2max}$ = 100 and $Z^*$ = 12 e). Adapted under terms of the CC BY-NC-ND 4.0 license [57], Copyright 2025.

We can now employ quantum chemistry to explore changes of bonding upon crystallization. As early as 1932 Zachariasen made the conjecture that oxide glasses possess the same short-range order in their vitreous form as in their crystalline state [58]. He even speculated why this would be the case and suggested that the forces between the atoms should be the same in the amorphous and crystalline state, explaining why the short-range order does not change upon crystallization. In the language of modern chemistry, one would argue that the chemical bonds between the atoms should be the same for both the vitreous and the crystalline phase. Quantum chemistry can prove or falsify this hypothesis. Interestingly, **Figure 5** confirms the hypothesis of Zachariasen from 1932. Indeed, the atomic arrangement and the chemical bonding descriptors for $SiO_2$ in the glassy and crystalline state are very similar, as shown in figure 5 for the nearest neighbor distance and the number of electrons shared between atoms [57]. Interestingly, this is not the case for GeTe, where both the nearest neighbor distance and the number of electrons shared between adjacent atoms change significantly upon crystallization. One can hence argue that GeTe forms a non-Zachariasen glass, i.e., a glass which differs significantly from its crystalline counterpart in terms of short-range order (and properties). These conclusions confirm the intimate link between changes of properties and changes in bonding upon crystallization.

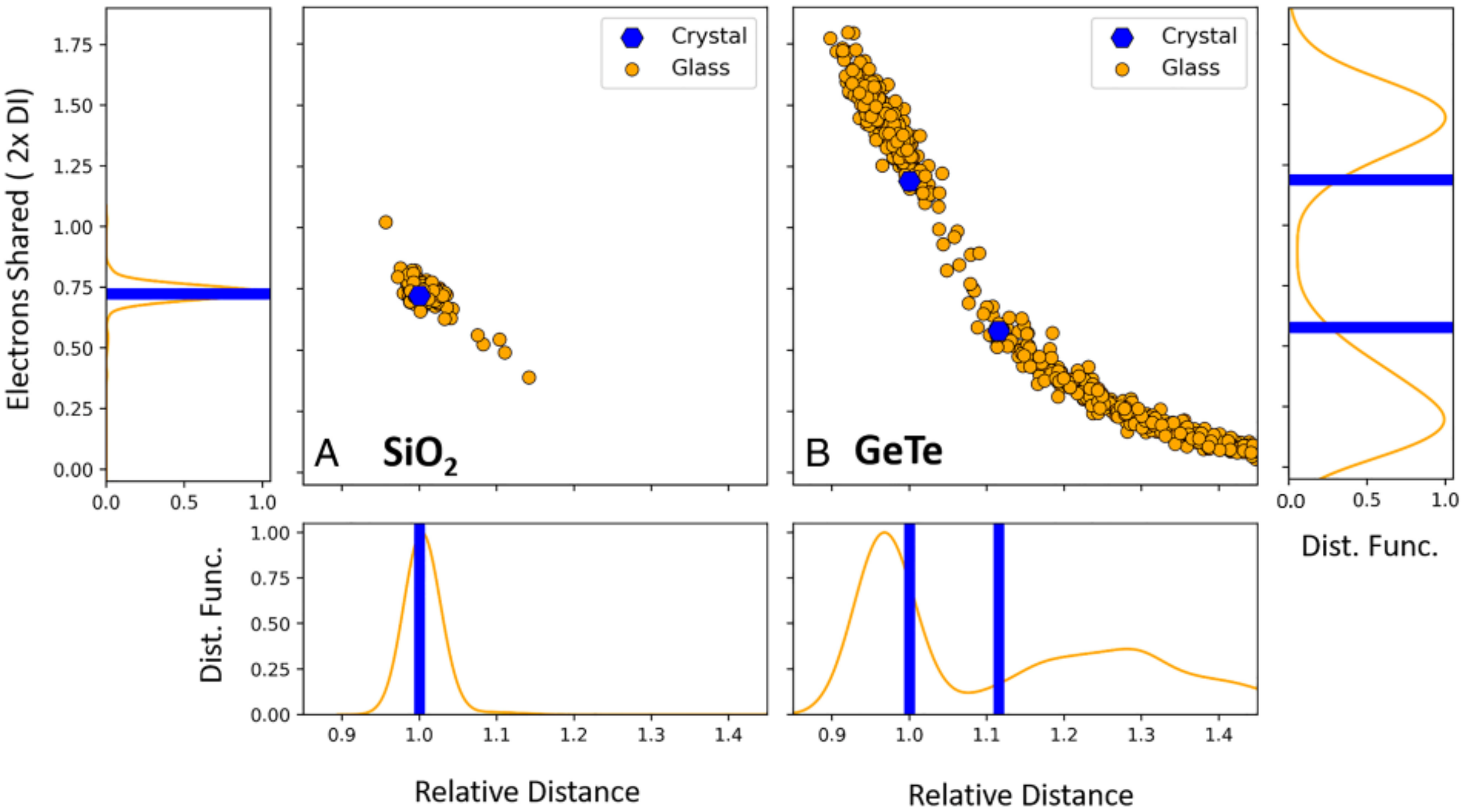

**Figure 5: Comparison of chemical bonding and atomic arrangement in glassy and crystalline $SiO_2$ (A) and GeTe (B).** For $SiO_2$, the atomic arrangement in the crystal (indicated by bars) and the glass (yellow line) is very similar; the glass only shows a small change in bond length compared to the crystal (Zachariasen glass). Interestingly, the quantum chemical bonding descriptor, the number of electrons shared between adjacent atoms is very similar for both phases, too. This is fundamentally different for GeTe, where the distribution functions for both the Ge-Te distances as well as the number of electrons shared between adjacent atoms differ significantly between the two phases. GeTe hence forms a non-Zachariasen glass. The relative distance is defined as the interatomic distance divided by the first neighbor distance in the crystal. Redrawn under terms of the CC BY-NC-ND 4.0 license [57], Copyright 2025.

This finding raises an interesting question: how can we identify solids which change their bonding upon crystallization? **Figure 6** answers this question. In this figure the change of bonding upon crystallization / vitrification is shown for six different solids. For $GeSe_2$, GeSe and $SiO_2$ only very small changes in bonding are observed. This is in line with Zachariasen's conjecture that the similarity of the properties and the atomic arrangement in the amorphous and crystalline state is due to pronounced similarities in bonding (Zachariasen glasses). On the contrary, only metavalent (crystalline) solids show a pronounced change of bonding upon vitrification; i.e., they form non-Zachariasen glasses. This explains why these materials change their properties so drastically upon crystallization [57].

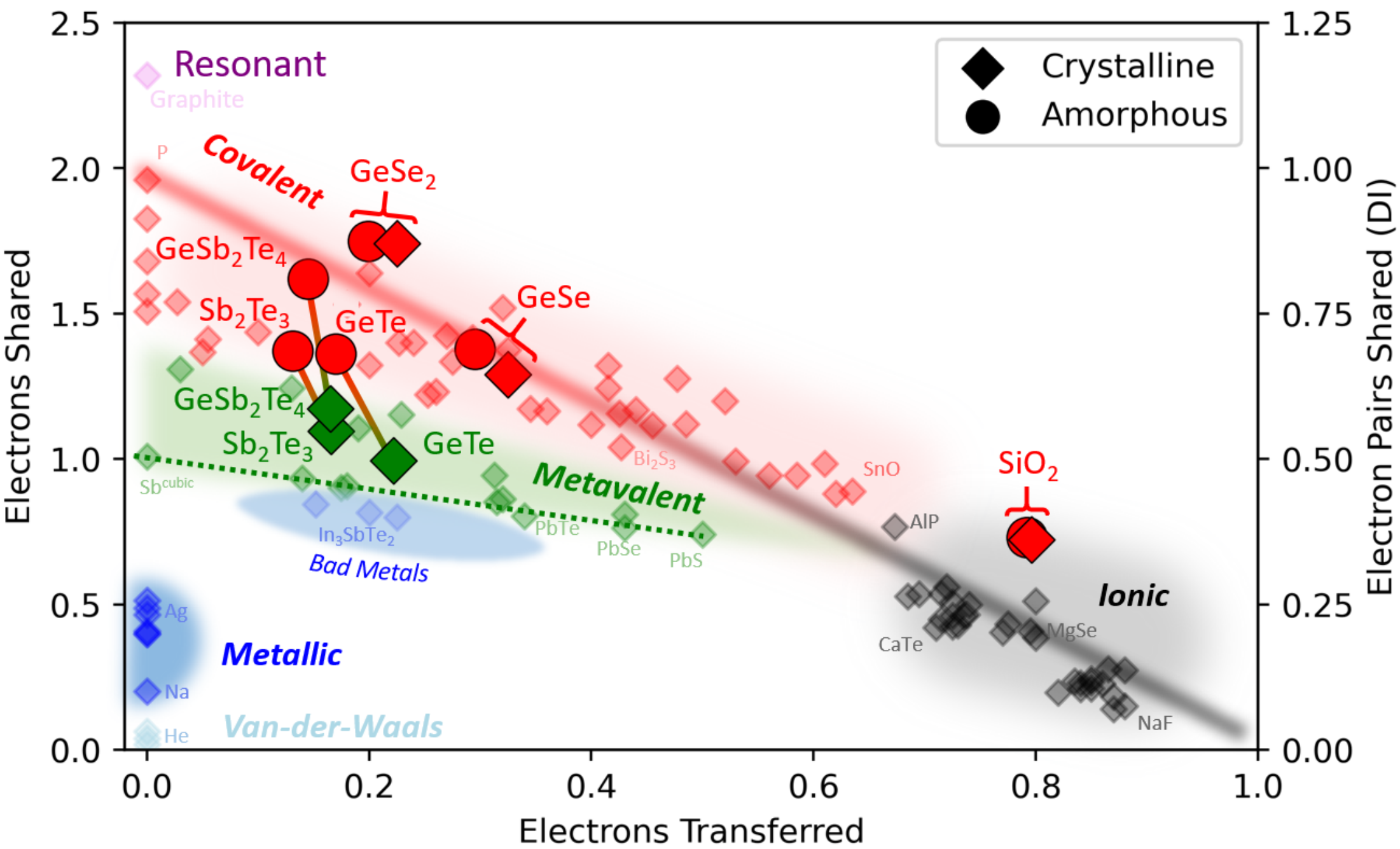

**Figure 6: 2D map classifying chemical bonding in crystals and glasses.** The map is spanned by the number of electrons shared (left y-axis) between adjacent atoms and the electron transfer renormalized by the formal oxidation state (x-axis). Different colors characterize different material properties and have been related to different types of bonds [42,43]. The glasses of three solids are characterized by a bonding mechanism which closely resembles the crystal ($GeSe_2$, $SiO_2$, and GeSe). On the contrary, for GeTe, $Sb_2Te_3$, and $GeSb_2Te_4$, pronounced changes in bonding occur upon crystallization. While crystalline GeTe, $Sb_2Te_3$, and $GeSb_2Te_4$ employ metavalent bonding, their glasses are covalently bonded. Metavalent crystals show characteristic features of quantum materials, i.e., they show a pronounced change of properties upon external stimuli like pressure or temperature. These materials change their bonding mechanism upon vitrification, while this is not the case for any other bonding mechanism in solids. Redrawn under terms of the CC BY-NC-ND 4.0 license [57], Copyright 2025.

There are several points to check if we have obtained a consistent picture for the working principle of phase change materials. Figure 6 implies that the vitrification of covalent solids like GeSe and metavalent solids like GeTe differs significantly. If this is the case, this immediately raises the question, how the transition from covalent to metavalent bonding looks like. To this end we can compare the behavior of pseudo-binary compounds between GeTe and GeSe and investigate how their properties depend upon composition. This is shown in Figure 7, where a number of compounds along the pseudo-binary line between GeTe and GeSe are compared in terms of their opto-electronic properties, represented here by the optical dielectric constant $\varepsilon_\infty$ and the bond rupture, represented by the PME.

Looking at the two border cases, GeTe and GeSe, fundamental differences can be seen. For GeTe, a pronounced difference of the optical dielectric constant $\varepsilon_\infty$ between the crystalline and the amorphous state is observed, which is not found for GeSe, where only a very modest change upon crystallization is observed [52]. The small change of $\varepsilon_\infty$ observed for GeSe can be attributed to a modest increase of density upon crystallization, which leads to a moderate increase of $\varepsilon_\infty$. This small change is predicted by the Clausius-Mosotti relation which links the density of a solid and its refractive index [59]. Concerning the bond rupture, it hardly changes upon crystallization in GeSe. This explains its modest property change. For GeTe, on the contrary, we find a pronounced change of bond rupture, which is clear evidence for a pronounced change of bonding from metavalent (crystalline) to covalent (amorphous). This strong change of bonding explains the pronounced change in optical properties.

We can now verify, if the transition from metavalent bonding in crystalline GeTe to covalent bonding in crystalline GeSe is gradual or if there is a pronounced change of properties upon changing composition. The former would question the idea of a clear border, while the existence of a pronounced step would favor the presence of two rather different bonding mechanisms. The data in **Figure 7** show for the crystalline phases a pronounced step for both $\varepsilon_\infty$ and the PME upon the transition from metavalent to covalent bonding. This confirms the hypothesis that in PCMs the pronounced change of optical properties upon crystallization is closely related to a change in chemical bonding. It furthermore demonstrates that crystalline PCMs employ a very peculiar bonding mechanism, defined by their unique property portfolio and an unconventional bond rupture in atom probe tomography [52]. Furthermore, they possess a well-defined property border to covalent bonding as well as a well-defined property border to metallic bonding (not shown here). Finally, they occupy a clear region in a quantum-chemical map, which separates the different bonding mechanisms, i.e., ionic, metallic, covalent and metavalent bonding. In this map, metavalent bonding is identified as a bonding mechanism which is not too ionic, i.e., there is not so much charge transfer between adjacent atoms and roughly one electron (i.e., half of an electron pair) is shared between adjacent atoms.

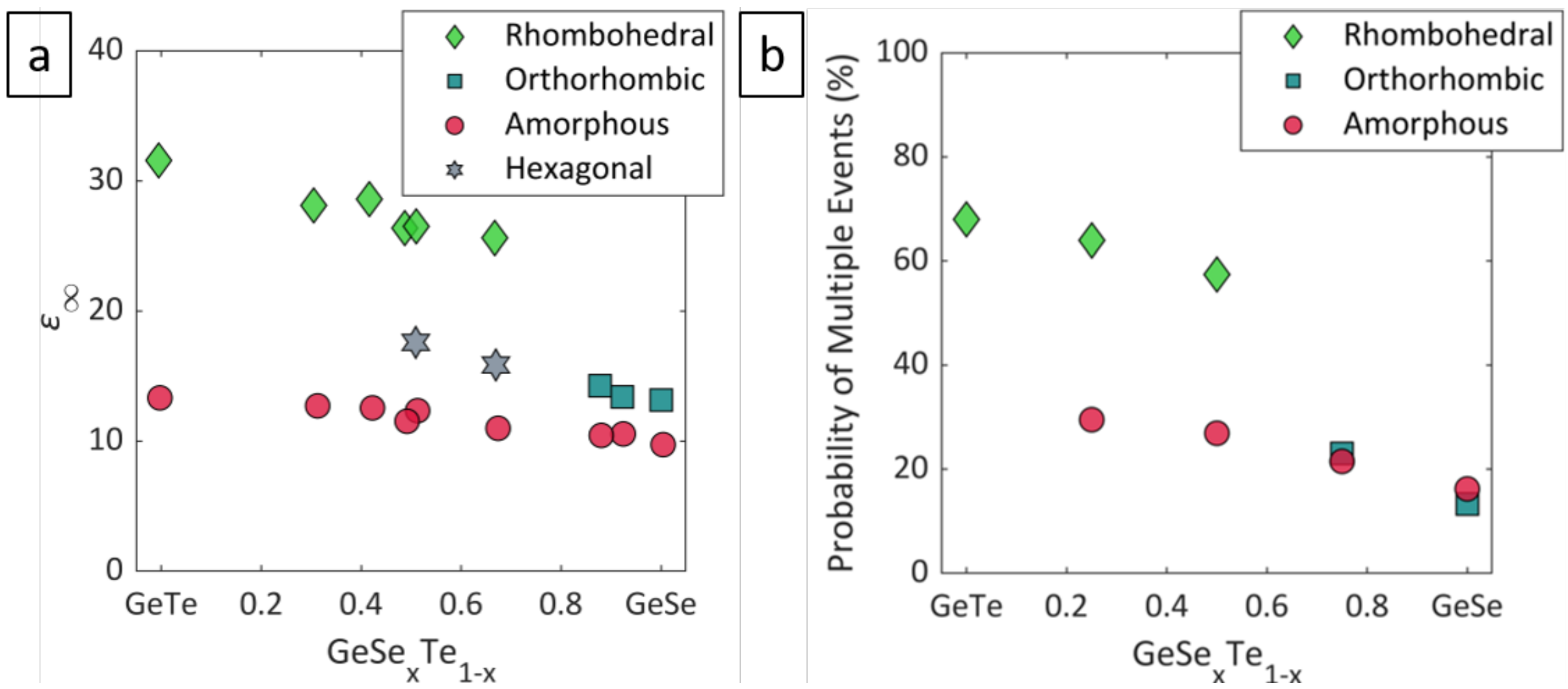

**Figure 7: Optical dielectric constant $\varepsilon_\infty$ and the probability of multiple events (PME) along the pseudo-binary line between GeTe and GeSe.** a) $\varepsilon_\infty$ as a function of stoichiometry. The rhombohedral phase, found up to 70% Se, is characterized by large values of $\varepsilon_\infty$, which exceed the value of the corresponding amorphous phases by more than 100%. Between 50% and 70% Se content the rhombohedral phase is metastable and transforms into a hexagonal phase upon further heating. This transition is accompanied by a pronounced drop in $\varepsilon_\infty$. b) Atom Probe Tomography confirms the transition from metavalent to covalent bonding between x = 0.5 and x = 0.75 for the crystalline solids. No major change of PME is found for the amorphous phases. Since atom probe tomography can quantify the bond rupture, characterized by PMI and PME, the changes of the optical properties are apparently connected to a change in chemical bonding [52]. The PME values were extracted from Ref. [44].

In the following, three crucial questions will subsequently be addressed:

a) Can we develop a deeper understanding of the nature of metavalent bonds and the resulting material properties?
b) Are there competing concepts to explain the property portfolio of PCMs?
c) How can systematic trends for chemical bonding be utilized to tailor PCMs?

Before doing so, it is important to add one point. Changes in optical properties upon crystallization can also be accomplished without a change of bonding. This can be seen for $SiO_2$, for example. The change in optical properties observed in $SiO_2$ upon crystallization is very small, since the corresponding density change of $SiO_2$ is very small, yet it is not zero. Materials with larger density change upon crystallization, such as $Sb_2Se_3$ or GeSe, thus can also show a significant change of the optical dielectric constant. However, this change is not related to a change in bonding, but just a change in density, described by the Clausius – Mossotti law and is typically much smaller [52].

## Can we develop a deeper understanding of the nature of metavalent bonds and the resulting material properties?

In the following we will focus on those solids which employ metavalent bonding in the crystalline state. Since metavalent bonding is classified by a plethora of unconventional features, one can ponder if it is possible to characterize and explain this bonding mechanism further. This can be done as shown in **Figure 8**, where one characteristic of the bond rupture of solids, the PME is shown as a function of the room temperature conductivity of a considerable number of binary compounds and elements. In this figure metals (depicted in blue) are on the right, since they have room temperature conductivities above $10^4$ S/cm. The majority of covalent solids possess much smaller electrical conductivities well below 10 S/cm. Like metals, they are characterized by low PME values of less than 30%. As already noted above, metals and covalent solids can be distinguished by their PMI, which is vanishing for metals, but non-vanishing for their covalent counterparts. Strikingly, much higher PME values (> 35 %), are observed in the transition zone between electron localization and de-localization. These findings indicate that during the transition from localized to delocalized electronic states a well-defined transition zone exists where solids exhibit a unique bond rupture.

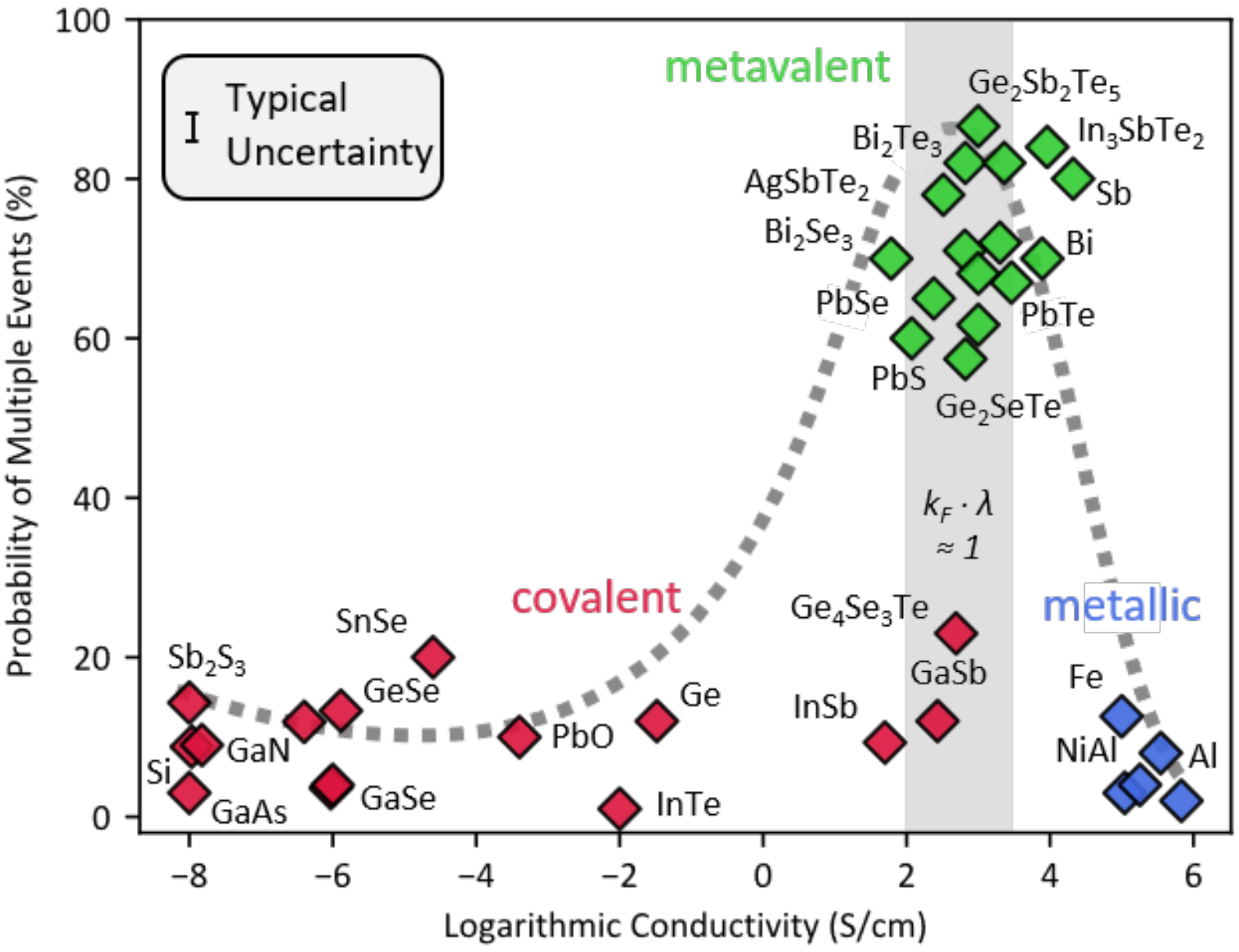

**Figure 8: Bond rupture (PME) for crystalline solids as a function of the electrical conductivity at room temperature.** While metals and covalent compounds possess a low PME of typically less than 25%, this is not the case for metavalent solids, which have a much higher PME. Their unusual bond rupture is found in the transition zone between localized and delocalized electrons, marked by $k_F \cdot \lambda \approx 1$. A fully labeled version of the figure is provided in the supplement. The typical uncertainty is displayed and is determined from the probability that an

evaporated multiple event is detected as a single event. Calculation and data are further given in the supplement. Data for most solids are taken from previous studies [53].

The transition from electron localization to delocalization can be quantitatively described using the product $k_F \cdot \lambda$, where $k_F$ is the Fermi wave vector and λ is the mean free path, according to the Mott-Ioffe-Regel criterion [60, 61]. Electron delocalization becomes significant when $k_F \cdot \lambda$ exceeds unity. This boundary delineates a narrow crossover region highlighted in Figures 8 and 9 and is further defined in the SI (Figure S2). While metals are located above the cross-over region, i.e., they are characterized by delocalized electrons, the majority of covalent solids are located below the cross-over region, and are thus characterized by localized electrons. Only metavalent solids, i.e., crystalline PCMs like GeTe, $Sb_2Te_3$, $In_3SbTe_2$ and $Ge_2Sb_2Te_5$ as well as related crystalline solids like $Bi_2Se_3$ are exclusively located in the grey transition zone.

This finding immediately raises two questions: Where are amorphous PCMs located? How does the bond rupture of solids look like that have conductivities in the cross-over region but do not employ metavalent bonds? The answer to both questions is found in **Figure 9a** and **9b**, respectively. Amorphous phase change materials have a much lower conductivity than their crystalline counterparts, as can be seen for $Ge_2Sb_2Te_5$, $In_3SbTe_2$, $GeSe_{.25}Te_{0.75}$ and $GeSe_{0.5}Te_{0.5}$. These amorphous solids all have conductivities, which are characteristic for solids with localized electrons. They show a PME typical for covalent compounds, in line with their property portfolio. This confirms that covalent solids can be identified by their characteristic bond rupture (nonvanishing PMI and low PME).

To verify that the bond rupture in atom probe tomography is indeed governed by the underlying bonding mechanism and not simply determined by the electrical conductivity, an important test has to be performed. How do the bonds break in APT, if the solid does not show the characteristic property portfolio of metavalent solids but has an electrical conductivity in the cross-over region? Such solids can be created, if a covalent solid like GaAs is doped. In Figure 9.b a series of GaAs samples has been studied with different levels of Si doping (doping concentration $n_{Si} = 2 \cdot 10^{16}$, $2 \cdot 10^{17}$ and $2 \cdot 10^{18}$ $cm^{-3}$). These solids have electrical conductivities of (11.8, 118 and 833 $Scm^{-1}$ [62]), i.e., two of these samples are located right in the cross-over region and one is located right below. Hence, we would expect high PME values, if only the electrical conductivity governs the bond rupture. This is not the case, showing that these doped semiconductors have a bond rupture which is characteristic for a covalent solid, in line with their property portfolio. This confirms that only metavalent solids show the peculiar bond rupture depicted in Figures 8 and 9. We are presently working on an atomistic model that explains this behavior.

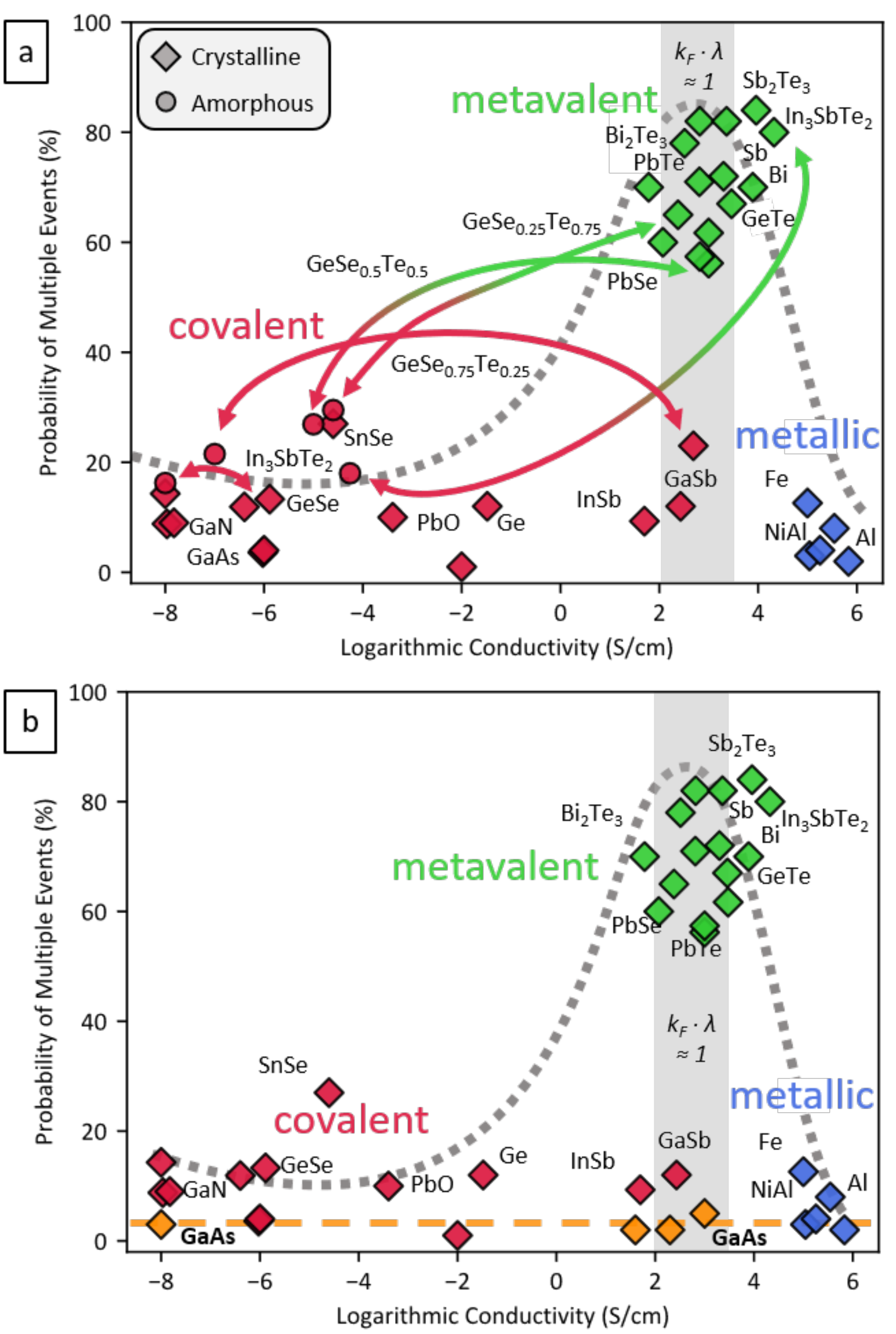

**Figure 9: Bond rupture (PME) for amorphous and crystalline solids as well as doped GaAs as a function of the electrical conductivity at room temperature.** a) Upon crystallization a number of amorphous chalcogenides ($In_3SbTe_2$, $GeSe_{0.75}Te_{0.25}$, $GeSe_{0.5}Te_{0.5}$, $GeSe_{0.25}Te_{0.75}$ and GeSe) change the bond rupture drastically ($In_3SbTe_2$, $GeSe_{0.5}Te_{0.5}$, $GeSe_{0.25}Te_{0.75}$), while others only show marginal changes (GeSe and $GeSe_{0.75}Te_{0.25}$). Those chalcogenides which change their bond rupture a lot all show the characteristic features of metavalent solids in their crystalline state. Electrical conductivities values are taken from [63]. b) The crystalline GaAs samples (yellow diamonds) span a wide range of conductivities, due to different levels of doping. Yet, their bond rupture remains unchanged and remains characteristic for covalent bonding. The unusual bond rupture shown by metavalent solids is thus not governed by the conductivities of the corresponding samples, lying in the grey transition region, but their unusual bonding. A fully labeled version of the figure is provided in the supplement.

Finally, we would like to mention another interesting aspect that can be derived from Figure 8. This is the question of the transition from metavalent bonding to metallic bonding, i.e., increasing electron delocalization as well as the transition to covalent bonding, i.e., increasing electron localization. Intriguing data have already been obtained for the transition to metallic bonding upon exploring topological chiral semimetals like AlPd, AlPt and related compounds. These solids have electrical conductivities slightly above the cross-over region, but have a PME higher than ordinary metals like Fe, NiAl and Al [64]. This implies that studying the transition from metavalent to metallic bonding can help to identify how electrons delocalize. Answering this question can provide insights onto a possible trajectory from localized to delocalized states, which should also help to unravel different paths from insulators to metals, i.e., MITs. Similar studies are presently also undertaken to understand the transition from metavalent to covalent bonding [65].

## Are there competing concepts to explain the property portfolio of PCMs?

As already pointed out several times, the pronounced change of optical and electronic properties upon crystallization is one of the hallmarks of PCMs. Hence, several attempts have been made in the last 50 years to develop a concept that can reproduce this pronounced change of optoelectronic properties. In solid-state physics, often the band structure is consulted to explain optical properties. However, we are not aware of any successful example to do so for PCMs. Apparently, knowing the band structure of such chalcogenides has not yet helped to identify a compound as a promising PCM.

Therefore, scientists have been pondering about alternatives to understand, explain and predict which solids show the unconventional property portfolio of PCMs, including the high Born effective charge Z*, the large value of the optical dielectric constant $\varepsilon_\infty$ and the pronounced anharmonicity, which leads to a high value of the transverse optical mode Grüneisen parameter $\gamma_{TO}$. Early on, scientists have been speculating that bonding is relevant to explain these unique properties. The first such attempt, which we are aware of, has been published by Lucovsky and White in 1973 [36]. The focus of their paper entitled 'Effects of Resonance Bonding on the Properties of Crystalline and Amorphous Semiconductors' has been the attempt to explain the unusual properties mentioned above, in particular the high values of Z* and $\varepsilon_\infty$, found in monochalcogenides such as PbTe, PbSe or GeTe. They argued that these properties could be attributed to resonance bonding, a concept devised by Linus Pauling [38]. They even argued that due to the lack of long-range order, resonance bonding would not be possible in the amorphous state. Hence, their bonding scenario could explain several crucial properties. Subsequent work has indeed shown that the majority of PCMs is characterized by pronounced changes of the dielectric function upon crystallization, including the optical dielectric constant $\varepsilon_\infty$ and the band gap, as well as the Born effective charge Z* [39]. These peculiar properties support the idea of an unconventional bonding mechanism. In line with the designation suggested by Lucovsky and White, these properties have been attributed to resonance bonding [39]. Nevertheless, it turned out later, that this assignment creates serious difficulties. Crystalline phase change materials like GeTe or other monochalcogenides such as SnTe and PbTe, possess very soft optical phonons indicative for a lattice instability. They also feature large Grüneisen parameters, a sign of anharmonicity and high values of the optical dielectric constant $\varepsilon_\infty$, a measure of the electronic susceptibility. Since these properties, which are determined by their bonding, are quantitatively and qualitatively different from "resonantly" bonded materials such as graphite, graphene or benzene, we conclude that the underlying fundamental mechanism must be different as well [42]. The designation of the bonds in these chalcogenides as "resonant" should thus be abandoned. Instead, we have suggested calling these materials 'incipient metals' and their bonding as 'metavalent'. The phrase 'incipient metals' has been proposed to stress that some of their properties, such as the electrical conductivity are almost metal-like. The word 'metavalent' has been invented to emphasize that the bonding is beyond, i.e., transcending, ordinary covalent bonding, therefore the Greek prefix 'meta' was chosen. At the same time, the word metavalent has also been chosen to reflect the nature of the bond between **meta**llic and co**valent**.

In the following, we will sketch further arguments showing that bonding in crystalline phase change materials is very different from bonding in benzene, graphene or graphite. In this review, APT has been employed to classify bonding in solids, showing that crystalline PCMs possess a rather unconventional bond rupture characterized by a large PME. Graphene nanotubes, on the contrary, do not show a high PME [66]. This supports the view that bonding in graphite and graphene is quite different from bonding in crystalline PCMs. The reason for this difference can be understood quite well. As shown in Figure 1, a resonantly bonded solid like graphite is located in a very different region of the map than the metavalent solids. While the latter are identified by sharing about 1 electron between adjacent atoms, i.e., about half of an electron pair, graphite is characterized by an Electrons Shared (ES) value above 2.3, i.e., higher than expected for a simple covalent bond (ES $\approx$ 2). This high ES value is attributed to the coexistence of two different bonds. In graphite, a $\sigma$-bond formed by the orbital overlap of $sp^2$-orbitals of adjacent atoms as well as a $\pi$-bond formed by the orbital overlap of adjacent $p_z$-orbitals occupied by just half of an electron pair prevail. The coexistence of a strong, covalent bond **and** a weak bond that resembles a metavalent bond elucidates the high ES values and the properties of this solid. In crystalline PCMs, on the contrary, only a single, weak bond with an ES around 1 exists, which is prone to lattice instabilities like Peierls distortions. This explains many of the characteristic properties of metavalent solids.

Interestingly, in recent years, further bonding mechanisms have been suggested to account for the properties of crystalline PCMs which challenge metavalent bonding [67, 68]. Before we discuss those, a number of crucial questions is listed which all those bonding mechanisms should be able to address: Can the bonding be quantified and depicted in a map like Figure 1? Can the corresponding scheme explain the relevant material properties? Can the bonding scheme explain the difference between crystalline and amorphous GeTe, and the lack thereof for GeSe? Finally, can the bonding mechanism describe the existence of a border between metavalent solids like GeTe and covalent bonding such as GeSe?

Two competing explanations for bonding in crystalline PCMs have been introduced recently that try to challenge the concept of metavalent bonding [67, 68]. In '*The myth of metavalency in phase change materials*', the authors argue that bonding in crystalline PCMs can be explained by the established concept of electron-rich, "hypervalent" bonds. These authors express the concern that a density-based approach, used to obtain Figure 1 does not adequately describe bonding in solids [69] [70]. If this concern is correct, the map shown in Figure 1 should be flawed. Fortunately, by now, both density-based and orbital-based calculations are available [45-48] that can both characterize bonds in solids. In **Figure 10**, the results of these two different calculation schemes are depicted.

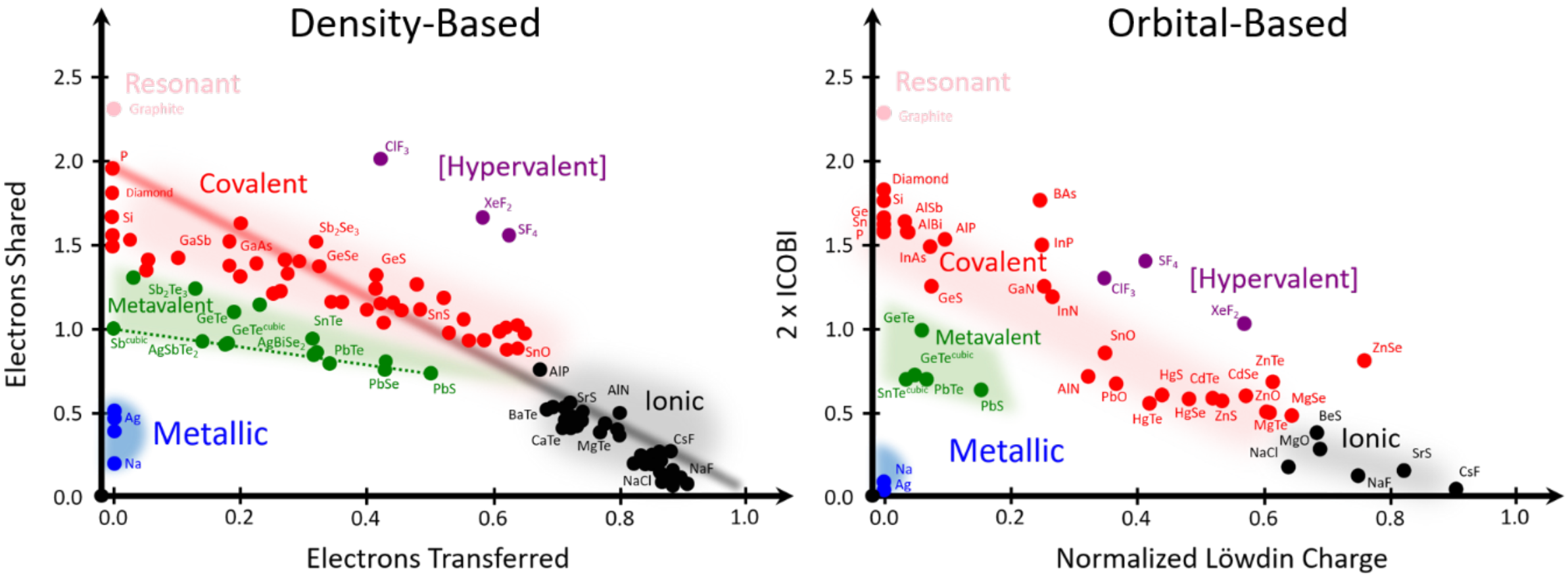

**Figure 10: 2D maps classifying chemical bonding in solids.** The map on the left is obtained from a density-based approach, while the map on the right is determined from an orbital-based approach. The x- and y-axis is spanned by the number of electrons shared between adjacent atoms and the electron transfer renormalized by the formal oxidation state (density-based calculation), or twice the bond order (ICOBI) as well as the Löwdin charge renormalized by the formal oxidation state (orbital-based calculation), respectively. Different colors characterize different material properties and have been related to different types of bonds [42] [43]. With the two different approaches, classes of solids (covalent, ionic, metallic and metavalent) are located in similar locations in both maps. This indicates that both approaches provide a consistent description of chemical bonds. Electron-rich compounds, such as $XeF_2$ or $SF_4$, frequently considered to be hypervalent, and metavalent solids, like GeTe or $Sb_2Te_3$, are located in different regions of the map. This is indicative for significant differences in their electronic states in the vicinity of the Fermi level, which govern bonding. The dashed green line in the density-based calculations identifies the location of solids with perfect octahedral arrangement like cubic Sb, $AgSbTe_2$ and PbS. Reprinted under terms of the CC BY 4.0 license [71], Copyright 2025.

It is comforting that in both computation schemes metavalent solids are located in the same area of the map, i.e., between the covalent and metallic region. This is an important conclusion since it shows that the program packages available [45-48] to quantify bonding in solids provide a coherent description of bonding. Furthermore, the region where crystalline solids like GeTe and $Sb_2Te_3$ are located in these maps, i.e., between metallic and covalent bonding, is the same in all three competing schemes, i.e., no matter if the bonding is called metavalent [42], hypervalent [67] or electron-deficient multi-center bonding [68]. The different labels apparently agree in the most important aspect, the quantification of the number of electrons transferred and electrons shared between adjacent atoms that was suggested in 2019 [49]. Apparently these two quantities are excellent quantum-chemical bonding descriptors. It is hence safe to conclude that by now the quantification of bonding in solids can be performed reliably and precisely. These maps also indicate that metavalent bonding is located in the competition zone between metallic and covalent bonding, i.e., between electron localization and electron delocalization. This conclusion is in line with the insights obtained by atom probe tomography depicted in Figure 8.

Given the general agreement for the two crucial bonding descriptors (ES and ET) for the three proposed bonding schemes, the question arises as to what the differences between these concepts are. This can be best seen if the three bonding names are replaced by their characteristic electronic bonding pattern. Metavalent bonding has been identified as a bonding scheme where two atoms are held together by a single electron (half of an electron pair) forming a 2 center – 1 electron (2c – 1e) bond rather than an ordinary 2c – 2e (covalent) bond. Since in crystalline PCMs, p- orbitals form the relevant bonds, these orbitals usually extend over adjacent atoms on both sides, i.e., forming 3c – 2e bonds. Hypervalent bonding, instead, is defined as 3c – 4e bonding, with the molecule $XeF_2$, being a characteristic representative. This bonding is electron-rich, while MVB is electron-deficient. Electron-deficient molecules, such as $B_2H_6$, which employ 3c – 2e bonding, are located in the same region as metavalent solids. This confirms the electron-deficient nature of metavalent solids, in line with further recent calculations [68]. Hypervalent molecules, such as $XeF_2$ or $SF_4$, on the contrary, lie above the covalent region and thus cannot be electron-deficient. Such 3c-4e bonds are well-known in molecules, but cannot prevail in solids in infinite chains. Further arguments, why the bonding in crystalline PCMs cannot be electron-rich have been summarized in [71]. Here we would like to add one comment. In a recent publication claiming that crystalline phase change materials utilize hypervalent bonding, the authors argue that this bonding mechanism can explain an important facet of these solids, i.e., their bond rupture in atom probe tomography [72]. In particular, they claim that larger clusters are released in APT for crystalline PCMs, which they consider as evidence for multicenter bonding. In **Figure 11** we show the probability of molecular ions (Figure 11a) as well as the mean fragment size (MFS) of these clusters (Figure 11b) as a function of electrical conductivity. This figure shows compelling evidence that both the PMI and the MFS are closely related to the electrical conductivity and **not** the bonding mechanism. Once samples are less conducting, i.e., have an electrical conductivity below $10^5$ S/cm, they show a non-vanishing PMI. This finding is linked to the penetration depth of the electrical field and not related to the bonding mechanism. For both covalent and metavalent solids, a non-vanishing PMI is found as well as an MFS above 1. This happens, once the electrical field can enter beneath the surface. Yet, the covalent compounds utilize ordinary two center – two electron (2c – 2e) bonding. Hence, hypervalent bonding is apparently unable to explain the characteristic bond rupture in metavalent solids. The task is not to explain the PMI, but the PME. Specifically, figures 3 and 9.b provide the true challenge. Which bonding mechanism can explain the pronounced change of bond rupture upon crystallization for those solids that we call metavalent? How can the pronounced maximum of the PME be explained for those peculiar solids between electron localization and electron delocalization?

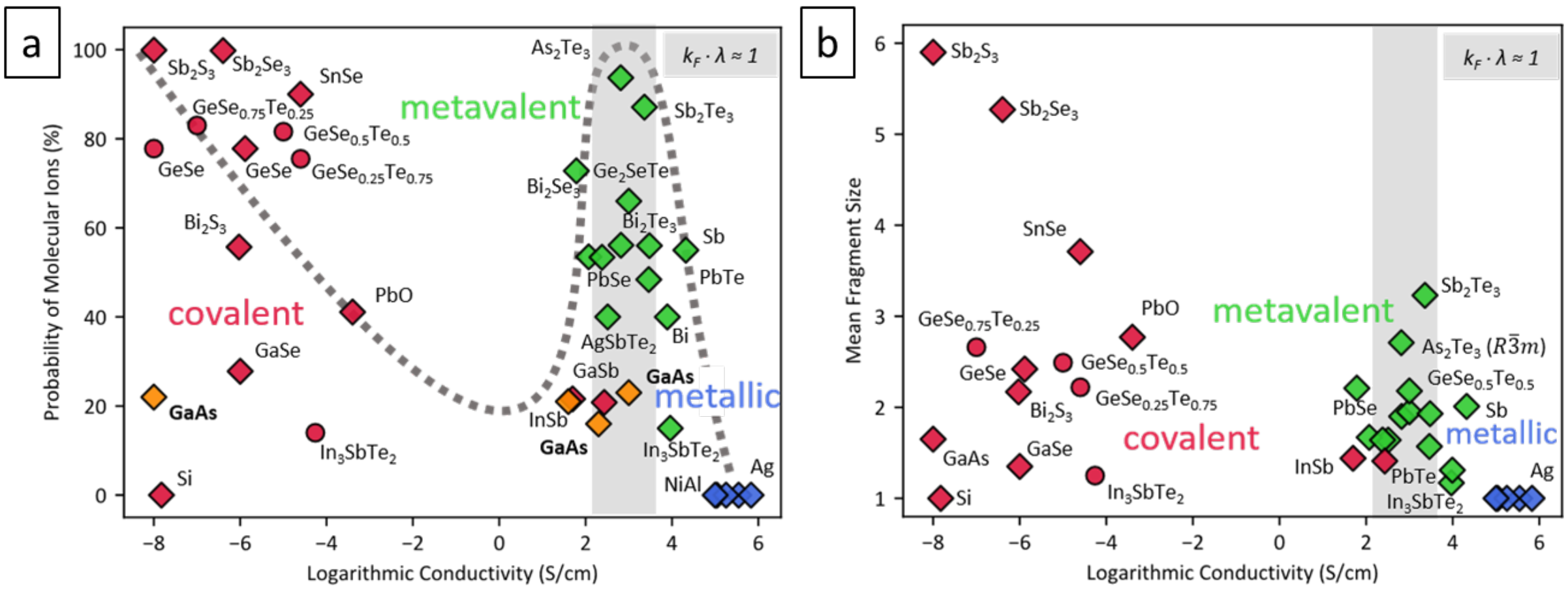

**Figure 11: Probability of Molecular Ions (PMI) and main fragment size (MFS) as a function of the electrical conductivity.** Once a critical conductivity of about $10^5$ S/cm is exceeded, the electrical field can no longer penetrate beneath the surface and molecular fragments are no longer dislodged (PMI = 0). Both metavalent and covalent solids possess a non-vanishing PMI. Covalent compounds display on average a higher MFS than metavalent compounds. Metals display an MFS of 1, as no molecular ions are dislodged. The evaporation behavior of GaAs remains constant independent of the dopant concentration (Figure 11.a).

Finally, Manon and coworkers argue that the 2c -1e bonding in crystalline PCMs should be described as electron-deficient multicenter bonding [68]. They claim that the concept of metavalent bonding ignores the multicenter character of the bonds. However, there has never been a doubt that the bonds in metavalent solids have a multicenter nature, i.e., that the 2c – 1e scenario can also be described as a 3c – 2e scenario as already shown and mentioned in [39, 50]. Electron-deficient (multi-center) bonds are also well-known in molecules and have already been identified many decades ago, e.g., in $B_2H_6$ molecules [73, 74], work that was later awarded by a Nobel prize for W. Lipscomb. Yet, it is consensus in chemistry, that these electron-deficient molecules employing 3c – 2e bonds form a subset of covalent molecules, which cannot be easily distinguished from ordinary covalent molecules based on their properties. Metavalent solids, on the contrary, can be easily distinguished from covalent solids by their properties and the difference in their bond rupture. Furthermore, metavalent solids have properties such as high values of the Born effective charge, pronounced optical absorption and intermediate electrical conductivities between good metals and undoped iono-covalent solids, which are not found in electron-deficient molecules. Hence, it seems misleading to utilize the same wording for bonds in different materials which differ in many relevant properties. To emphasize this aspect, we have proposed the name metavalent for the corresponding bond in solids.

Chemical bonding is possibly the most central concept in chemistry. In order to develop its full potential, it is important to develop bonding concepts that are capable of explaining key material properties [75]. Hence, we believe it is very important to develop bonding concepts that can help to classify the properties of materials. This requires a terminology which can differentiate solids like crystalline GeTe, PbTe or $Sb_2Te_3$ from compounds which employ metallic or covalent bonding. As can be seen in Figure 3, atom probe tomography can distinguish these three classes of materials based on differences in their PMEs and PMIs. Based on the unique set of properties of metavalent solids we are convinced that it is justified to give the bond in these materials a distinct name, different from terms used for other systems such as molecules that do not show a similar property portfolio. After all, the distinct bond rupture provides experimental evidence for a distinct and rather unusual bonding pattern. Hence, we have suggested six years ago to coin the 2c – 1e bonding (3c – 2e bonding) in solids metavalent bonding to distinguish it from electron-deficient, i.e., 3c – 2e bonding in molecules.

## How can systematic trends for bonding be utilized to tailor PCMs?

In this final chapter we raise one last question: How can the concepts reviewed here be employed to predict material properties? The map shown in Figure 6 already provides a first important answer. If we look for materials with a pronounced contrast of opto-electronic properties, we need to search for crystalline phases which employ metavalent bonding, i.e., are located in the green region of Figure 6. Yet, this map can help us to proceed significantly further in designing material properties. Two related quantities relevant for applications are the band gap and the strength of the optical absorption. Figure 6 can be used to predict such trends as a function of the position of the material in the green region of the map, as shown for example in fig. 4 of [57] and explained there. If we move from hypothetical cubic Sb (ES=1, ET = 0), to solids with larger ET, the charge transfer between the atoms opens up a band gap that increases with increasing ET. The band gap can also be increased, by increasing sizes of Peierls distortions, which increase the number of electrons shared in the shorter bond (leading to an increased ES value). Hence, ES and ET are not only good quantum-chemical bonding descriptors, they are also excellent property predictors. One can thus use this map to tailor material properties by moving to designated positions in the map upon alloying elements opening up an innovative material design strategy.

Interestingly, there are even property trends for certain quantities, where a systematic trend would not necessarily have been expected. Two such quantities are depicted in **Figure 12**, which displays the minimum crystallization time τ and the reduced onset temperature $T_{rg}$ for glass formation. Both quantities change systematically with their position in the map. This verifies that the concepts reviewed here enable a systematic property design of PCMs.

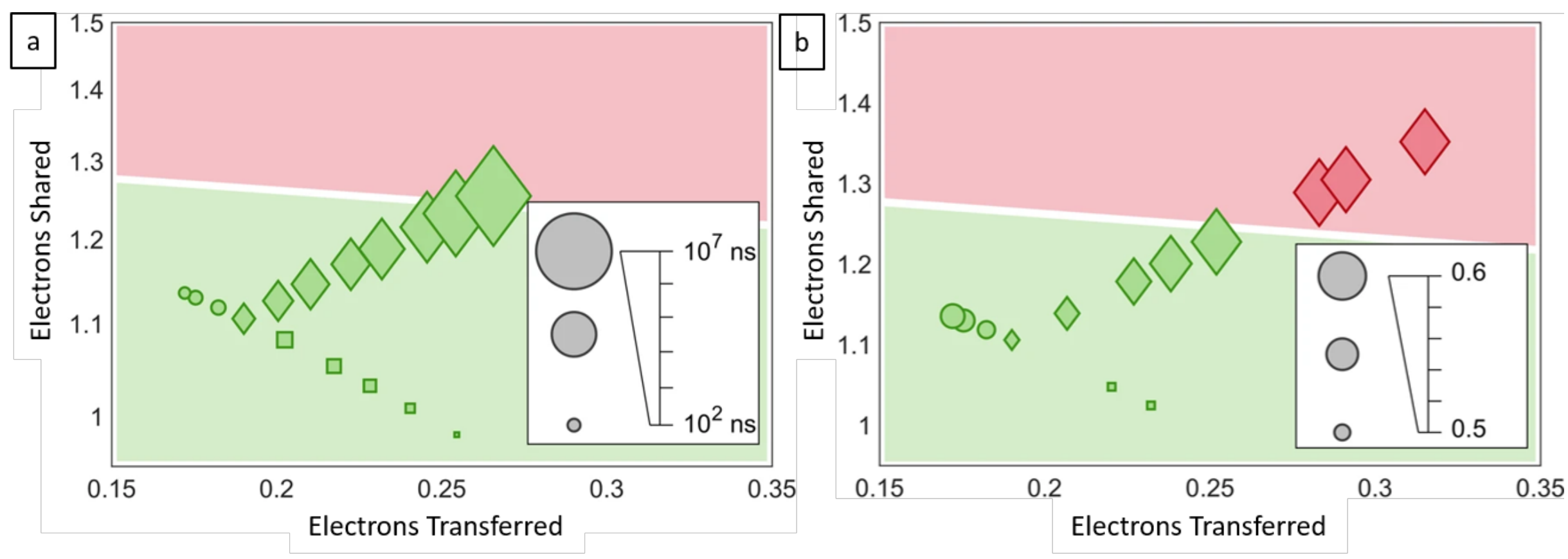

**Figure 12:** The minimum crystallization time τ (a) and the reduced onset temperature $T_{ro}$ for glass formation (b) displayed in relation to the electrons shared and electrons transferred for $GeTe_{1-x}Se_x$ (diamond), $Ge_{1-y}Sn_yTe$ (square) and $Ge_{1-z}Sb_{2z}Te_{1+2z}$ (circle). Both minimum crystallization time and the glass-forming ability decrease in the metavalent (green) region. Reprinted under terms of the CC BY 4.0 license [76], Copyright 2025.

Yet, there are a number of other possibilities to employ the concepts reviewed here. Take for example the different selenides that have been recently studied for photonic switches and similar optical applications in the visible range, where tellurides are usually too strongly absorbing. $Sb_2Se_3$, $Sb_2S_3$ and GeSe are such selenides that have recently been studied and explored, since they have a much lower absorption in the near IR (NIR) [77, 78]. How do they differ from phase change materials like GeTe and $Sb_2Te_3$? Atom probe tomography provides a clear answer. For GeSe and $Sb_2S_3$, we do not find a change of bond rupture upon crystallization. Both the amorphous and crystalline phase possesses a bond rupture typical for covalent solids, i.e., a rather low PME and non-vanishing PMI. In these materials, the change in optical properties is attributed exclusively to the increase in density as described by the Clausius-Mossotti relation. Tellurides like $Sb_2Te_3$ and GeTe, instead, undergo a change of bonding upon crystallization, where the crystalline phase employs metavalent bonding as seen from the large PME. For these solids, the change in optical properties is much larger than expected based on the density change alone. Hence, APT can be utilized to understand the contrast of optical properties upon crystallization.

Atom probe tomography can also be employed to identify more solids which utilize metavalent bonding. Recently, we have shown, e.g., that compounds like $AgBiSe_2$ also feature metavalent bonds [79]. This material is an example for a silver chalcogenide which has attractive thermoelectric properties. Apparently, there are many thermoelectrics with favorable properties which show a bond rupture characteristic for metavalent bonding. It will be interesting and presumably very rewarding to design thermoelectrics to understand why so many good thermoelectric materials utilize metavalent bonds. First attempts to unravel this have been focusing on the unique band structure of metavalent solids [80].

Furthermore, another exciting question can be tackled where atom probe tomography can play a prominent role. As can be seen from Figure 8, metavalent solids are located in the competition zone between electron localization, as in covalent bonding and electron delocalization, as in metallic bonding. Yet, in terms of the PME trends shown in Figure 8, there is a gap between these incipient metals (metavalent solids) and true metals. This raises the question if a distinct class of solids can be identified which is located between metavalent solids and metals. Indeed, there is mounting evidence that topological chiral semimetals form such a distinctive class [64].

Finally, it is tempting to study the limits of metavalent bonding upon increasing electron localization, i. e., at the border to covalent bonding. In particular, one can ponder if there are tellurides which do not (!) employ metavalent bonding. We have investigated and compared the properties of $Sb_2Te_3$ with $In_2Te_3$ [81, 82]. This comparison shows striking differences concerning properties, which indicates that $In_2Te_3$ employs ordinary covalent bonding. This conclusion helps to understand why thin films of $In_2Te_3$ do not show a pronounced change of properties with film thickness unlike films of metavalent sesqui-chalcogenides [81].

## Summary and Outlook

Recent advances in understanding the remarkable properties of phase change materials have been reviewed here. Atom probe tomography plays a prominent role in this endeavor, since it has demonstrated that the bond rupture of most phase change materials differs significantly between the amorphous and crystalline phase. The probability of multiple events (PME) i.e., the likelihood that more than one ion is dislodged per successful laser pulse in laser assisted field evaporation, is particularly important in this regard. The changes in bond rupture, i.e., the PME, provide compelling evidence that bonding changes significantly upon crystallization in phase change materials. Crystalline PCMs are characterized by a PME above 55%, not found for metals or iono-covalent solids. This confirms that crystalline PCMs employ a unique bonding mechanism coined metavalent bonding (MVB). While crystalline PCMs employ MVB, amorphous PCMs behave as covalent solids characterized by a much lower PME. PCMs thus change their bonding upon crystallization. Crystalline solids with a high PME lie in a narrow conductivity range between metals and iono-covalent solids, indicative for a competition between electron localization and delocalization. This finding is confirmed by quantum chemical calculations which show that two quantum chemical bonding descriptors can be utilized to quantify bonding in solids, the number of electrons transferred and the number of electrons shared. A map spanned by these two quantities shows that metavalent solids are located in a region where approximately one electron is shared between adjacent atoms and bonding is not too ionic. This quantum chemical bonding map is now used to find and explain property trends relevant for PCMs in various application domains.

The present review demonstrates unequivocally that atom probe tomography can be utilized to distinguish different bonding mechanisms, i.e., metallic, covalent and metavalent bonding. This is a remarkable finding considering how much controversy has been created to distinguish bonding mechanisms in solids. Atom probe tomography provides two crucial quantities to distinguish different bonding mechanisms, the probability to form molecular ions (PMI), as well as the probability to detect more than one ion from the tip surface upon laser-assisted field evaporation (PME). The PMI is vanishing for solids with metallic bonding only, since it is closely related to the field penetration depth. A high PME, on the contrary is only found for metavalent solids. This high PME is observed exclusively for solids in a narrow conductivity range which marks the competition zone between localization and delocalization. Yet, doped semiconductors which fall into the same conductivity range do not show this bond rupture. This observation should help to unravel the origin of the unusual bond rupture of metavalent solids. Work is in progress to identify the reason of the unconventional bond rupture of these compounds.

Furthermore, we note that the transition from metavalent to metallic and from metavalent to covalent solids should be studied and explored further. In particular, the transition to the metallic state should be insightful, since this change is closely related to a metal – insulator transition, more precisely, an insulator to metal transition. Initial evidence indicates that metavalent solids undergo an unconventional transition to the metallic state, where electron delocalization is intimately interwoven with an inherent lattice instability.

Finally, we have focused in this review on solids where a single bond determines the material properties. What happens in those solids where several different bonds play a significant role, as in Zintl phases, for example. Will APT help to unravel the bonding pattern in more complex solids, as well?

**Supporting Information**

Supporting Information is available from the Wiley Online Library or from the author.

**Author Contribution**

J.K. and M.W. have designed and developed this study and wrote the present publication with input from all co-authors.

**Acknowledgments**

The authors acknowledge financial support from NeuroSys as part of the initiative “Clusters4Future”, which is funded by the Federal Ministry of Research, Technology and Space BMFTR (03ZU2106BA) as well as support by the Deutsche Forschungsgemeinschaft within SFB 917 (Nanoswitches). The authors gratefully acknowledge computing time provided by the NHR Center NHR4CES at RWTH Aachen University (project number p0020357). The critical reading of the manuscript by Dasol Kim and Yuan Yu is gratefully acknowledged. The help of Felix Hoff in the revision of the manuscript is gratefully acknowledged. The fruitful discussions about the PME uncertainty with Thomas Schmidt and Niklas Penner are gratefully acknowledged.

**Conflict of Interest**: The authors declare no conflict of interest.

**Data Availability Statement**: The data that support the findings of this study are available from the corresponding author upon reasonable request.

# Supporting Information

# Understanding and Designing Phase Change Materials: Insights from Atom Probe Tomography

*Jan Köttgen[1], Nils von den Driesch[2], Alexander Pawlis[2], Matthias Wuttig[1,2]*

**J. Köttgen, Prof. M. Wuttig**

[1] Institute of Physics IA, RWTH Aachen University, 52074 Aachen, Germany

* E-Mail: wuttig@physik.rwth-aachen.de

**Dr. N. von den Driesch, Dr. A. Pawlis, Prof. M. Wuttig**

[2] Peter-Grünberg-Institute – JARA-Institute Energy Efficient Information Technology (PGI-10) Wilhelm-Johnen-Straße, 52428 Jülich, Germany

## Local Electrode Atom Probe Tomography

The Local Electrode Atom Probe (LEAP) comprises a local electrode, a pulsing near-UV laser, and a position-sensitive detector. The specimen's tip, with a diameter below 100 nm, is cooled to cryogenic temperatures (typically 30-60K) within an ultra-high vacuum chamber. Positioned approximately 10 µm from the tip-shaped specimen, the local electrode applies a static voltage in the range of a few thousand volts. To compensate for the increasing tip diameter with ongoing evaporation, the voltage rises continuously to apply a constant electric field. A voltage or laser pulse initiates the evaporation at specific time intervals; ions are then emitted and accelerated from the specimen surface towards the position-sensitive detector. The ions' time-of-flight between pulse initiation and detector impact correlates with their mass to charge ratio, facilitating the determination of chemical identities.

## Evaporation Descriptors

The field evaporation of different compounds can be described by different quantities. The Probability of Multiple Events (PME) is described as the ratio of the number of ions that are detected on the same pulse as another ion ($X_{ME}$) to the total number of detected ions consisting of detected ions from multiple events $X_{ME}$ and single events $X_{SE}$. On the latter, only one ion is detected during an evaporation event. Probability of Molecular Ions (PMI) is described as the ratio of detected molecular ions ($X_{MI}$) to the total number of detected ions consisting of the number of molecular ions $X_{MI}$ and single ions $X_{SI}$, consisting of only one atom.

## Mean Fragment Size

In addition to the two previously introduced bond rupture descriptors (PME and PMI), the mean fragment size (MFS) is used to further investigate the evaporation behavior. It is defined as the arithmetic mean of the number of atoms per fragment $a_i$ in regards to the total number of fragments $n_i$ Figure S1 gives an example for the analysis of the relevant quantities.

## Determination of the PME

The goal for the PME determination is an experiment with the lowest amount of uncorrelated evaporation and molecular dissociation. Therefore, each material is measured using an individual set of parameters, with a focus on optimizing the laser pulse energy. The laser pulse frequency used is typically rather low, as some of the metavalent and covalent materials evaporate with very large molecular ions, resulting in a high time-of-flight. Each experiment is analyzed by investigating the corresponding correlation histogram. If DC-evaporation (trails starting from the origin) are observed, the laser pulse energy is raised to increase the number of ions that evaporate immediately after the laser pulse impact. If the tails of the mass peaks are elongated, indicating a high tip surface temperature induced by a high laser pulse energy, the latter is reduced. Thus, the highest number of ions evaporates directly after the laser pulse reaches the tip. Through this approach the materials can be characterized with the required attention due to their very unique evaporation behaviors and allows the determination of a distinct set of PME, PMI and MFS values.

In the case shown in Figure S5 the PME is extracted from the experiment performed at 10 pJ, as this the lowest laser pulse energy that shows no DV-evaporation.

## LEAP comparison

*The used laser wavelengths of the different LEAP setups were reduced with ongoing development of the devices. Lee et al. showed a difference of the PME measured via the Invizo 6000 compared with the LEAP 5000. However, due to the very different setup designs, this difference can not only be attributed to the different laser wavelengths. The differently shaped local electrodes combined with the Invizo's dual laser setup and decelerating lenses make the influence of the wavelength hard to separate.*

*A more realistic comparison of the LEAP 5000 XR and the 6000 XR by Schiester et al. makes it obvious that the multiplicity histograms generated from the LEAP 5000 XR and 6000 XR are very similar.*

*Despite this evidence, we investigated the imaginary part of the dielectric function of the typical metavalent material GeTe as a function of the wavelength, shown in Figure S6: It shows that for higher wavelengths the imaginary part eps2 as a metric of the absorption increases. Thus, the laser energy can be absorbed even more for higher wavelengths, making the LEAP 3000, 4000 and 5000 the most suitable for the investigation of metavalent materials.*

*We can also consider another aspect of the varying available LEAP setups:*

*The detectors of the newest LEAP generation (LEAP 5000 & 6000) have an open detector area of 80%. When using a reflectron (LEAP 6000 XR and LEAP 5000 XR), the resulting efficiency drops to 52%. The previous generation of detectors (LEAP 3000 & 4000) had an open area of 55%, which is reduced to 37% with the use of a reflectron.*

*All data is gathered with a LEAP 5000 XS, meaning that the detector efficiency is as high as possible and thus causing the lowest uncertainty that a multiple event is actually observed as a multiple event. We will discuss this point in more detail in the answer of question #6.*

*Compare with https://www.sciencedirect.com/science/article/pii/S030439912500004X*

## PME uncertainty determination

Every evaporated ion is observed with a probability of ε on the detector, given by the open area of the detector and the time flight (ε = 80% for a LEAP 5000 XS).

Considering an example of a double multi-hit evaporation, this event is detected as a multiple event with a probability of $\varepsilon^2$, as a single event with $2(1-\varepsilon)\varepsilon$ and as no event with $(1-\varepsilon)^2$. For the case of a LEAP 5000 XS this corresponds to 64% for the detection of a double event, 32% of a single event and 4% for the detection of no event.

The probability of a multiple event of the Multiplicity M being detected as a single event (k=1) can be generalized with the binomial coefficient as

$$P(k=1|M) = \binom{M}{k}\, \varepsilon^k (1-\varepsilon)^{M-k}$$

For a given dataset with $\binom{M}{1} = M$ and the number of events i per Multiplicity $i_n$ we receive:

$$\Delta_{Multiples\ to\ Singles} = \sum_{n=1}^{M} n\, \varepsilon^1 (1-\varepsilon)^{n-1}\, i_n$$

Analogously, we receive

$$\Delta_{Multiples\ to\ Zero} = \sum_{n=1}^{M} (1-\varepsilon)^n\, i_n$$

and

$$\Delta_{Singles\ to\ Zero} = (1-\varepsilon) i_1$$

We can then describe the uncertainty on the multiple events as

$$\Delta_{ME} = \Delta_{Multiples\ to\ Singles} + \Delta_{Multiples\ to\ Zero}$$

and the uncertainty on the single events as

$$\Delta_{SE} = -\,\Delta_{Multiples\ to\ Singles} + \Delta_{Singles\ to\ Zero}$$

For the final uncertainty of the PME we propagate

$$PME = \frac{X_{ME}}{X_{ME} + X_{SE}}$$

to

$$\sigma_{PME} = \frac{\sqrt{(X_{SE}\,\Delta_{ME})^2 + (X_{ME}\,\Delta_{SE})^2}}{(X_{SE}+X_{ME})^2}$$

The corresponding uncertainties for a variety of values are displayed in Table S1. A statistical analysis is performed via a boxplot analysis in Figure S7.

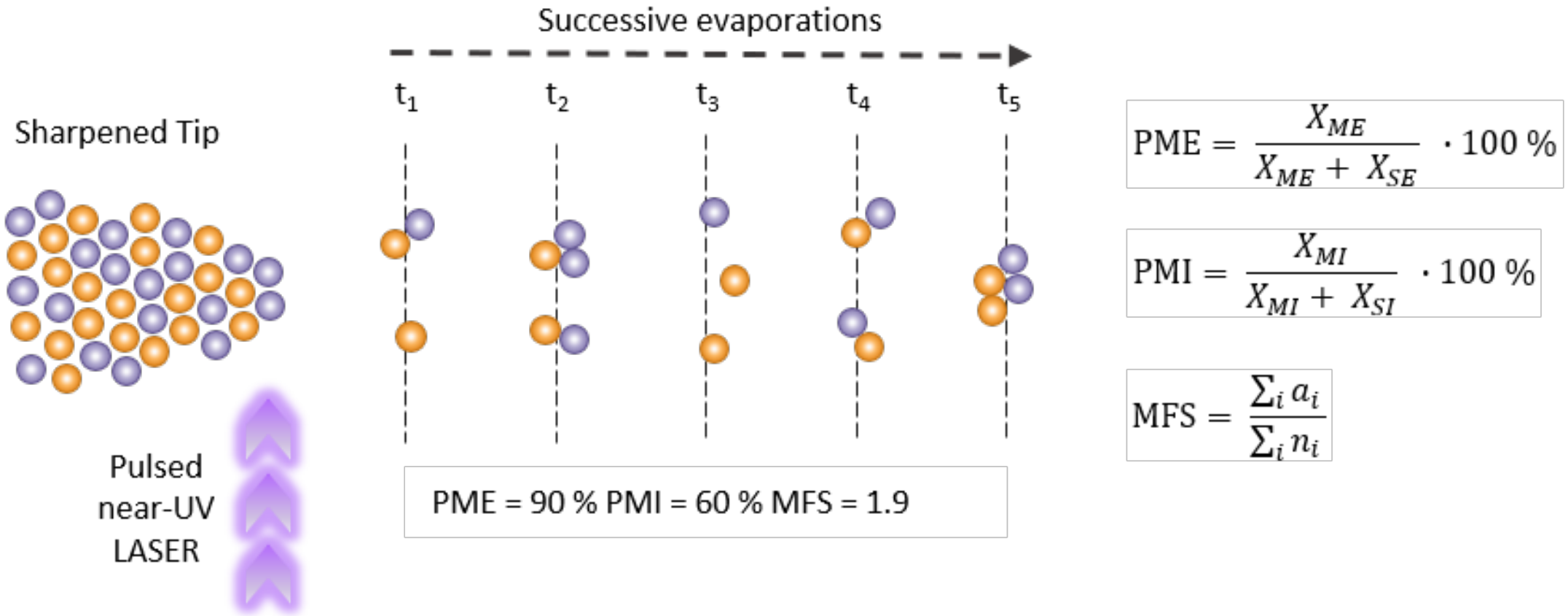

**Figure S1:** Schematic depiction of the used atom probe tomography analysis. PME, PMI and MFS are defined and explained for an example measurement.

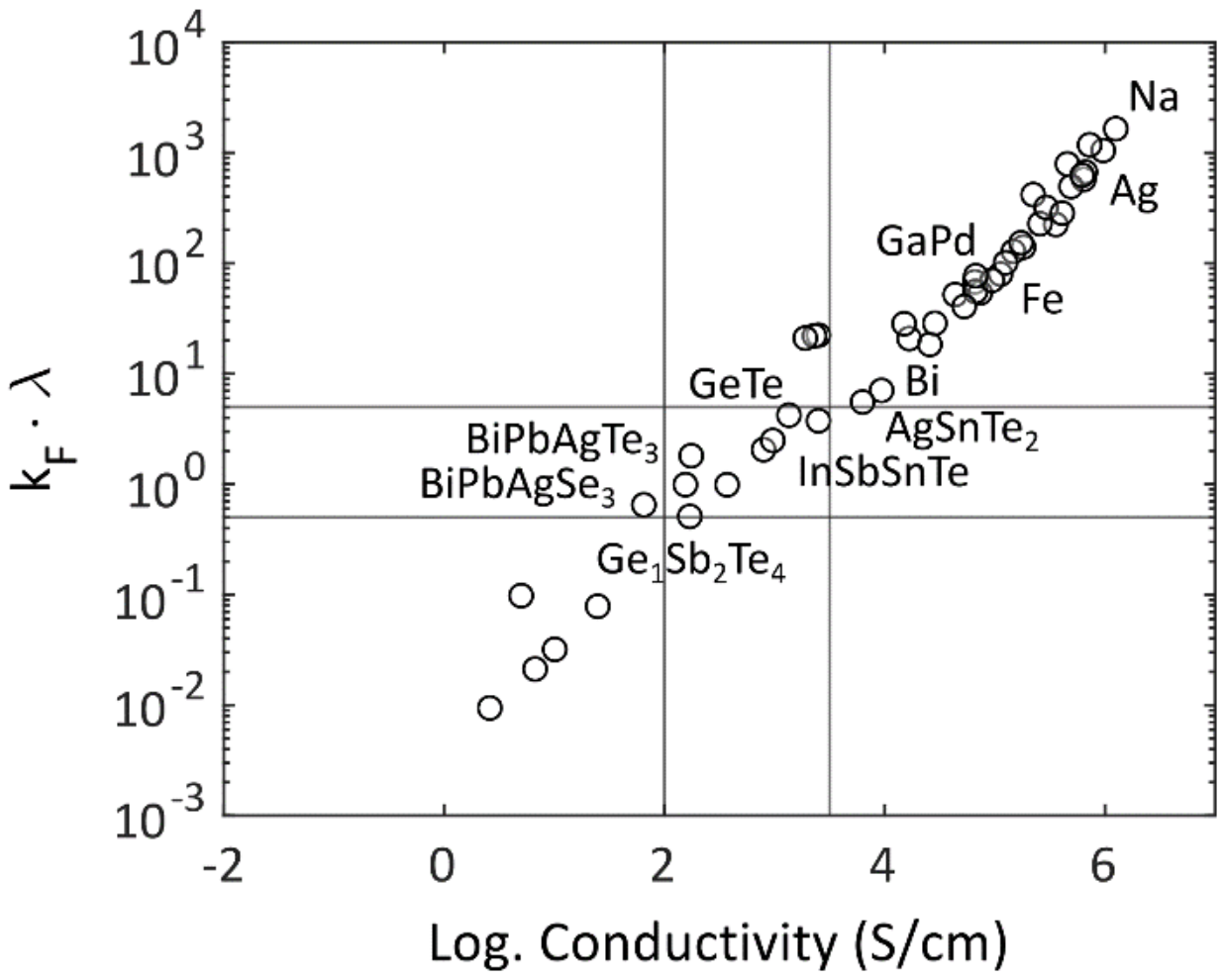

**Figure S2**: Dimensionless product $k_F \cdot \lambda$ versus DC conductivity for a wide range of materials. The parameter $k_F \cdot \lambda$ compares the electron mean free path ($\lambda$) to the Fermi wavelength ($1/k_F$), serving as an indicator of metallic transport. Solid lines around $k_F \cdot \lambda = 1$ mark the Mott-Ioffe-Regel limit, where metallic conduction gives way to localization as conductivity decreases. These data were used to define the gray bars in all main text figures.

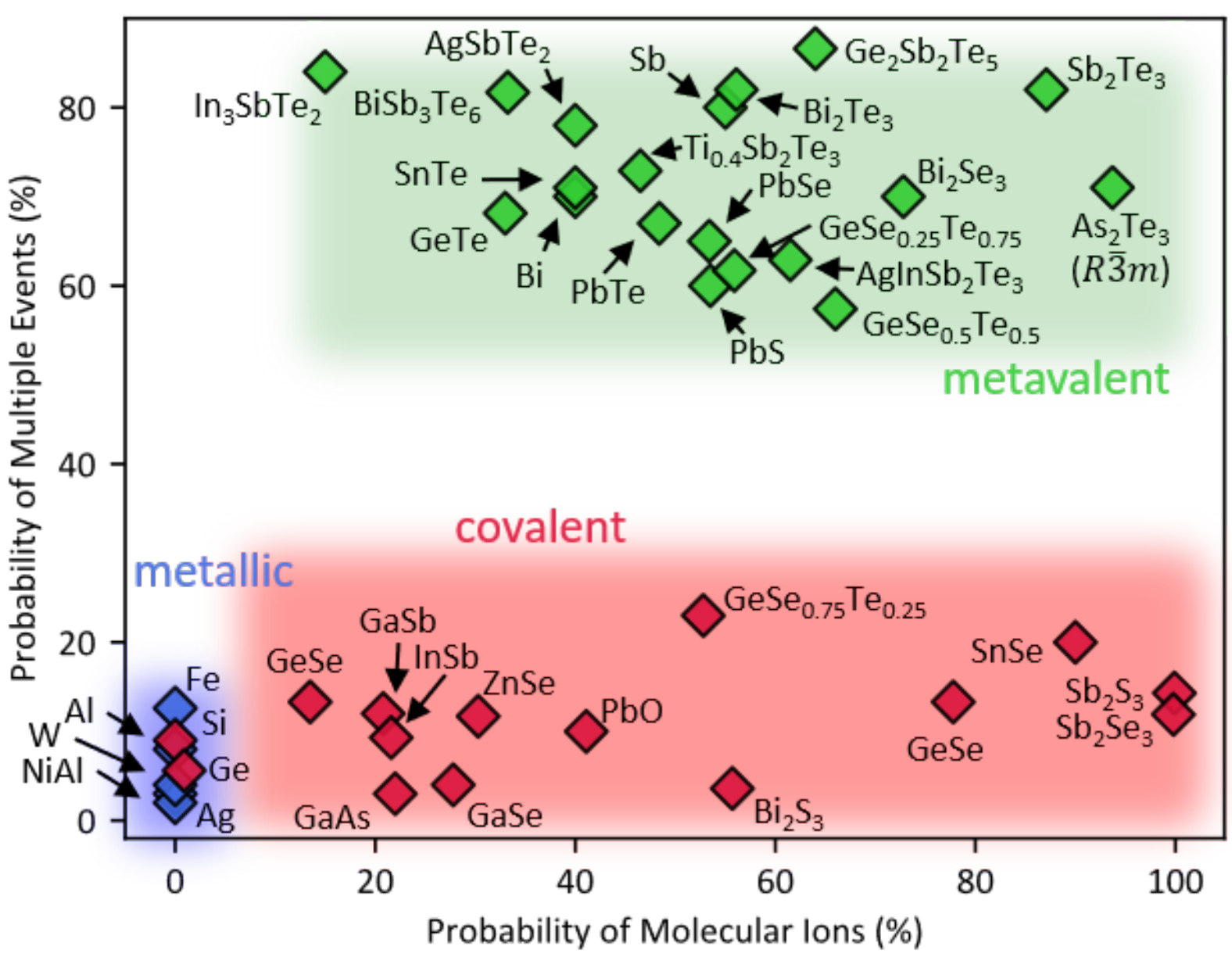

**Figure S3:** Fully labeled version of Figure 2 in the main text. Characteristic pattern of bond rupture for different crystalline solids.

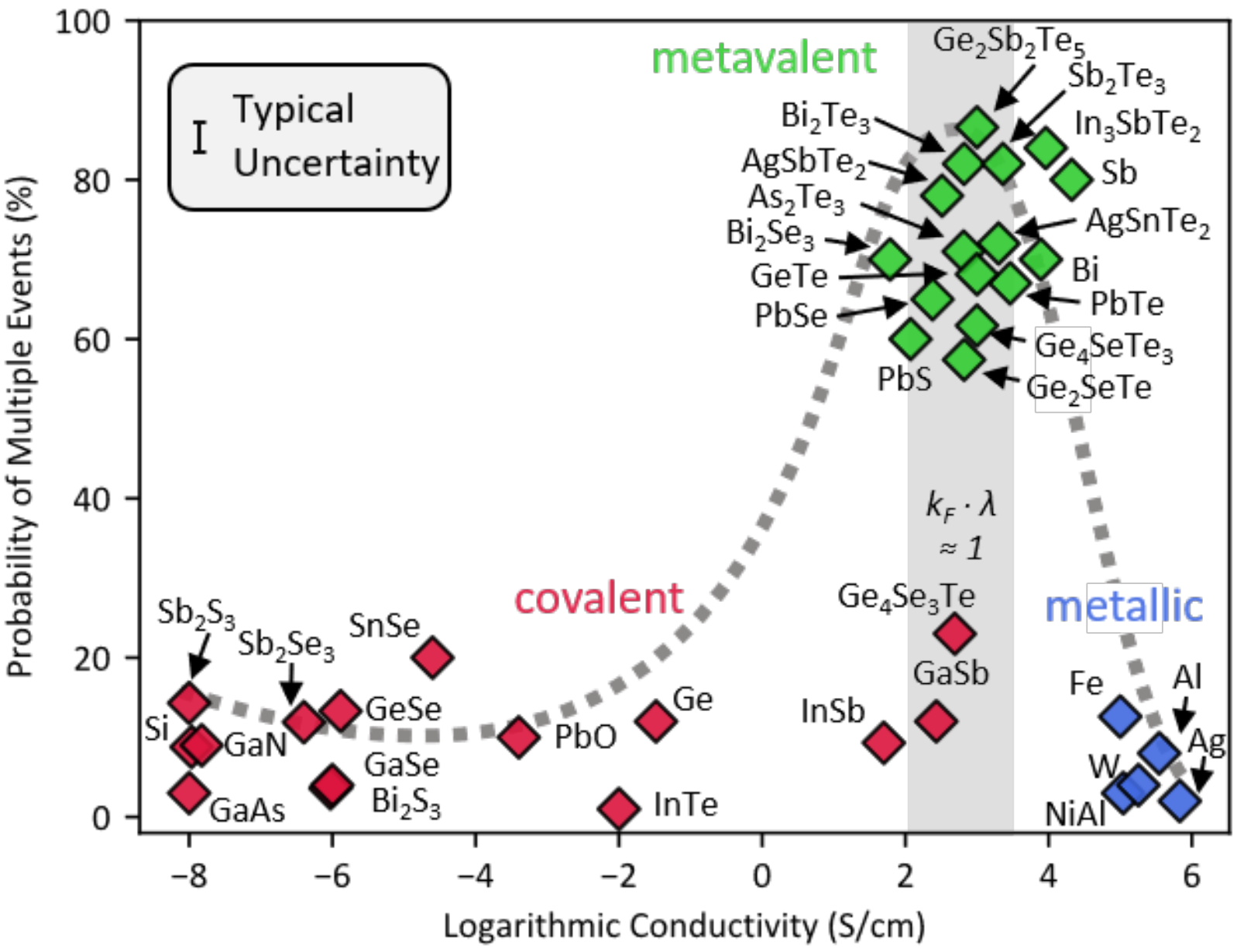

**Figure S4**: Fully labeled version of Figure 8 in the main text. Bond rupture (PME) for crystalline solids as a function of the electrical conductivity at room temperature. The typical uncertainty is displayed.

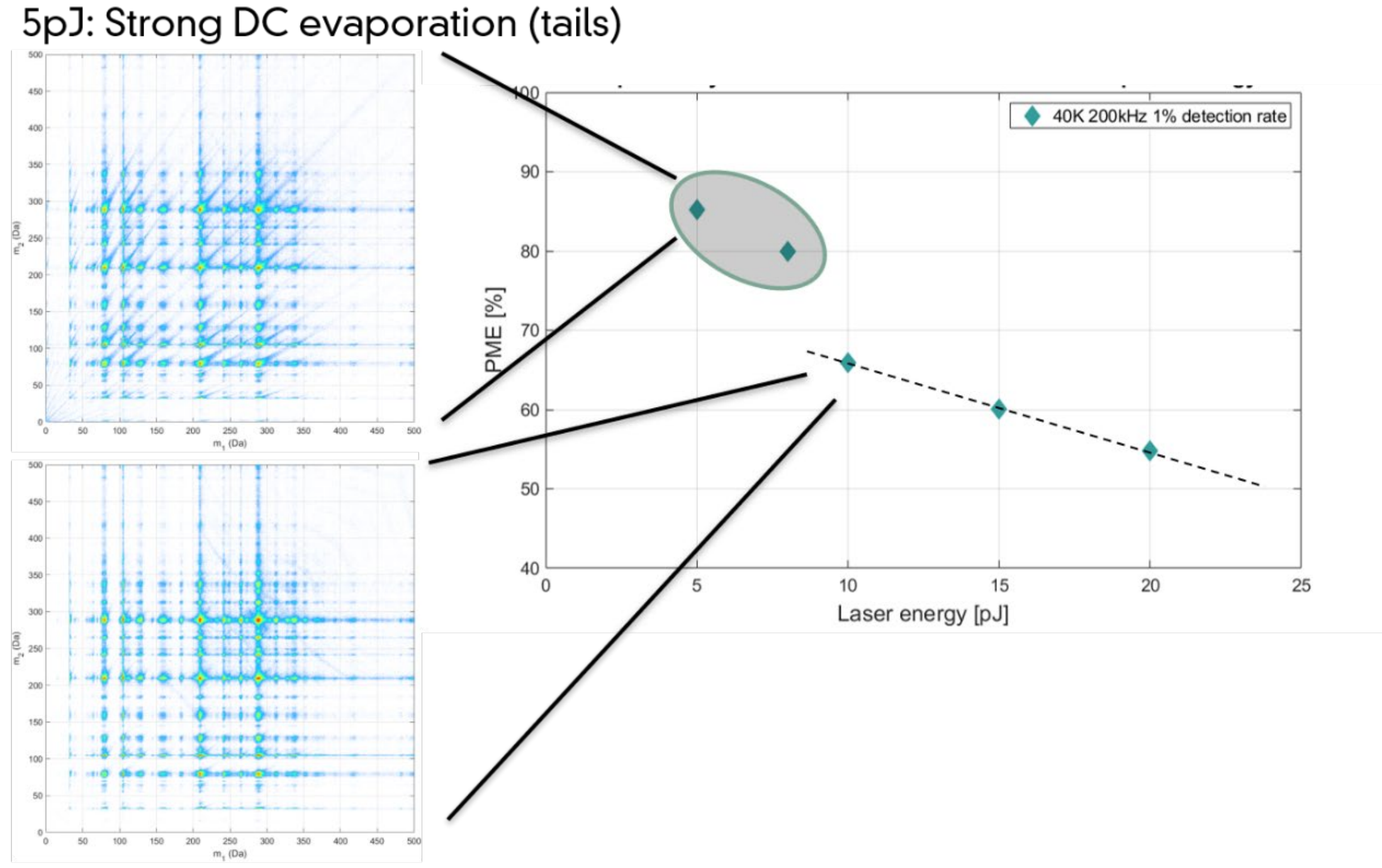

*Figure S5: The PME is determined by using an experiment with no DC-evaporation or dissociation in the corresponding correlation histogram.*

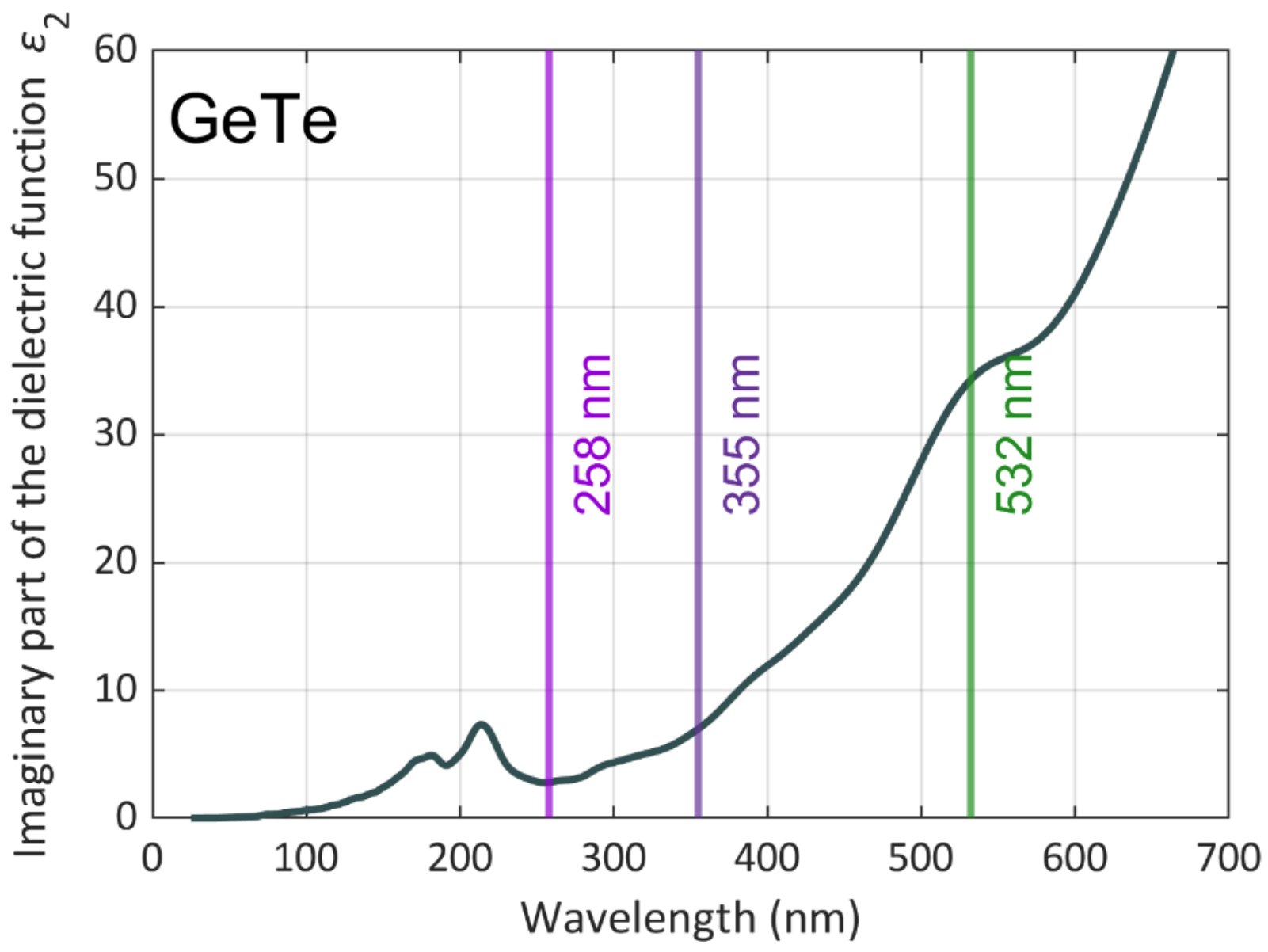

*Figure S6: The imaginary part of the dielectric function of the metavalent compound GeTe shown with the different laser wavelengths of the commercial LEAPs.*

| Sample | Error PME (%) | PME (%) |
| --- | --- | --- |
| AgBiSe2 | 3,7 | 58,3 |
| AgSnSe2 | 3,0 | 53,5 |
| AgSnTe2 | 2,5 | 69,8 |
| Al | 1,2 | 3,6 |
| As2Te3 | 1,9 | 75,0 |
| Bi2S3 | 1,1 | 3,4 |
| Bi2Se3 | 2,7 | 6,4 |
| Bi2Te3 | 1,2 | 8,1 |
| Bi | 6,4 | 59,0 |
| GST | 6,8 | 86,6 |
| GaSb | 2,8 | 10,4 |
| GaSe | 2,0 | 6,8 |
| GeSe | 4,2 | 19,8 |
| GeTe | 4,3 | 65,7 |
| IST | 9,6 | 84,4 |
| In2Se3 | 2,1 | 7,3 |
| InSb | 2,5 | 9,8 |
| InTe | 3,9 | 12,2 |
| PbO | 2,6 | 9,2 |
| PbS | 0,4 | 74,4 |
| PbSe | 0,5 | 70,8 |
| PbTe | 6,7 | 56,5 |
| Sb2S3 | 3,3 | 11,9 |
| Sb2Te3 | 3,0 | 67,3 |
| Sb | 1,3 | 79,0 |
| Sb | 0,2 | 75,9 |
| Si | 3,9 | 12,2 |
| SnSe | 4,4 | 18,8 |
| SnTe | 4,1 | 71,4 |
| Sn | 5,8 | 3,0 |

*Table S1: The uncertainties for the PME of different solids.*